\documentclass[prc,twocolumn,showpacs,amsmath,amssymb]{revtex4}
\usepackage{rotating}
\usepackage{amsmath,amssymb}
\usepackage{verbatim}
\usepackage{graphicx}
\usepackage{dcolumn}
\usepackage{amstext}
\usepackage{lscape}
\usepackage{mathptmx}
\usepackage{verbatim}
\usepackage[colorlinks,citecolor=blue,bookmarks]{hyperref}
\usepackage{times}

\begin{document}

\title{Quest for magicity in hypernuclei}
 
\author{M. Ikram$^{1}$, Bharat Kumar$^{2}$, S. K. Biswal$^{2}$ and S. K. Patra$^{2}$}
\affiliation{$^1$ Department of Physics, Aligarh Muslim University, Aligarh-202002, India\\ 
$^2$ Institute of Physics, Sachivalaya Marg, Bhubaneswar-751005, India.}

\email{ikram@iopb.res.in}

\date{\today}

\begin{abstract}
In present study, we search the lambda magic number in hypernuclei within 
the framework of relativistic mean field theory (RMF) with inclusion of 
hyperon-nucleon and hyperon-hyperon potentials. 
Based on one- and two-lambda separation energy and two-lambda shell 
gap; 2, 8, 14, 18, 20, 28, 34, 40, 50, 58, 68, 70 and 82 are suggested to be the 
$\Lambda$ magic number within the present approach. 
The weakening strength of $\Lambda$ spin-orbit interaction is responsible for emerging the 
new lambda shell closure other than the model scheme.
The predicted $\Lambda$ magic numbers are in remarkable agreement with 
earlier predictions and hypernuclear magicity quite resembles with nuclear magicity.
Our results are supported by nuclear magicity,
where neutron number N = 34 is experimentally observed as a magic which is one 
of the $\Lambda$ closed shell in our predictions. 
In addition, the stability of hypernuclei is also examined by calculating the binding energy 
per particle, where Ni hypernucleus is found to be most tightly bound triply 
magic system in considered hypernuclei.   
Nucleon and lambda density distributions are observed and it is found that 
introduced $\Lambda$'s have significant impact on total density and reduces 
the central depression of the core nucleus.
Nucleon and lambda mean field potentials and spin-orbit interaction 
potentials are also observed for predicted triply magic hypernuclei and the 
addition of $\Lambda$'s affect the both the potentials to a large extent.
The single-particle energy levels are also analyzed to explain the shell gaps for 
triply magic multi-$\Lambda$ hypernuclei.

\end{abstract}
\maketitle

\section{Introduction}\label{int}
The study of hypernuclei has been attracting great interest of 
nuclear physics community in providing the information from 
nucleon-nucleon (NN) interaction to hyperon-nucleon (YN) and 
hyperon-hyperon (YY) interactions. 
Due to the injection of hyperons a new dimension is added in normal nuclear 
system and hyperons serve as a potential probe for exploring many nuclear 
properties in domain of strangeness~\cite{bando1990,tan2004,lu2011}. 
However, hyperon-nucleon interaction is weaker than nucleon-nucleon 
but it is imperative as well as important to describe 
the nuclear many-body system with strangeness.
Various theoretical approaches Skyrme Hartree Fock~\cite{3,4,5,6,7,8,9,10,11}, 
relativistic mean field~\cite{14,15,16,17,18,19}, 
cluster, variational, diffusion Monte Carlo~\cite{Hiyama2010,Hiyama2012,Nemura2005,nemura2005,Gal2009,Gal2011,Usmani2006,Usmani2008,Zhou2008,Vidana2004,Samanta2008}, and G-matrix~\cite{25,26,27} have been 
employed by scientific community to estimate the strength of 
hyperon-nucleon as well hyperon-hyperon interactions. 
Further, these models have established themselves as very effective 
in testing the existence of bound hypernuclei and the stability of nucleonic 
core against hyperon(s) addition or the occurrence of exotic strange 
matter which facilitate the path toward multi-strange systems.

Magic numbers in nuclear physics are  certain neutron and proton numbers 
in atomic nuclei, in which higher stability in the ground state 
is observed than in the neighbouring nuclides and are most abundant in nature.
The various experimental signatures that show discontinuity at magic numbers 
are: the energy required for the separation of one and two nucleons, 
the energies of alpha and beta transitions, pairing energy and the 
excitation of low-lying vibrational~\cite{goeppert1958,solovev1992,nilsson1995}. 
The separation energy is sensitive to the collective or single particle 
inter play and provides a sufficient information about the nuclear 
structure effects. 
The discovery of magic numbers paved the way to great progress in 
understanding of nuclear structure and these numbers became 
the cornerstones for future theoretical developments in nuclear physics. 

It is worthy to mention that the several signatures are seen for the 
evolution of the magic gaps along the nuclear chart including superheavy 
region~\cite{sorlin2008,patra2012,zhang2005}. 
The quest for proton or neutron magic numbers in the elusive mass 
region of superheavy nuclei is of utmost importance as the mere 
existence of superheavy nuclei is the result of the interplay between the 
attractive nuclear force (shell effects) and the large disruptive 
coulomb repulsion between the protons that favours the 
fission~\cite{strutinsky1967,strutinsky1998}. 
It is well established that 2, 8, 20, 28, 50, 82 and 126 are the nucleonic magic number.
In addition to this, Z = 120 and N = 172, 184 are predicted to be next 
magic number by various theoretical models in 
superheavy mass region~\cite{patra1999,patra1997,patra2004,rutz1997,meng2007}.
These predictions have been made on the basis of separation energy, shell gaps, 
pairing energy and shell correction energy etc.
It may therefore relevant to extend the 
line of thought to the hypernuclear chart. 
It is well known that the spin-orbit interaction in $\Lambda$ channel is weaker 
than nucleonic sector and thus the $\Lambda$ magic numbers are expected to be 
close to the harmonic oscillator ones: 2, 8, 20, 40 and 70. 
In this paper, our main motive is to make an extensive investigation to search 
the $\Lambda$ magic number within the RMF approach and to obtain the 
stability of triply magic system with doubly magic core.

The magic numbers in nuclei are characterized by a large shell 
gap in single-particle energy levels. 
This means that the nucleon in the lower level has a comparatively 
large value of energy than that on higher level giving rise to more stability. 
The extra stability corresponding to certain number can be 
estimated from the sudden fall in the separation energy. 
The $\Lambda$ separation energy is considered to be one of the key quantity to 
reveal the nuclear response to the addition of lambda hyperon.
Therefore, in present work, we obtain the the binding energy per particle and 
one-lambda as well as two-lambda separation energies for considered multi-hypernuclei.
Moreover, two-lambda shell gaps is also calculated to make the clear presentation of 
the magic number which also support to the two-lambda separation energy.
To mark the $\Lambda$ shell gaps, single-particle energy levels are analyzed 
that may correspond to $\Lambda$ magic number.
In addition, to analyze the structural distribution as well as impact of 
$\Lambda$ hyperon on bubble structure for considered nuclei, 
total (nucleon plus $\Lambda$) density is reported.    
Nucleon and lambda mean field and spin-orbit interaction 
potentials are also observed.
On the basis of binding energy per particle, the stability of 
triply magic hypernuclei is reported.

RMF theory has been quite successful for studying 
the infinite nuclear systems and finite nuclei including the 
superheavy mass region~\cite{patra1999,patra1997,patra2004,rutz1997,meng2007,serot1992,gambhir1990,ring1996,serot1986,boguta1977,jha2007,ikram2013}.
It is quite successful to study the equation of state for infinite nuclear 
matter as well as pure neutron matter, where the existence of strange 
baryons is expected~\cite{glendenning1998,schaffner2002}.
In this context, addition of strangeness degree of freedom to RMF formalism is 
obvious for the suitable expansion of the model and such type of attempts have 
already been made~\cite{schaffner1994,rufa1990,glendenning1993,schaffner1993,mares1994,sugahara1994,vretenar1998,schaffner2002,lu2003,shen2006,win2008,ikram14}.
RMF explains not only the structural properties of singly strange hypernuclei but also 
provides the details study of multi-strange systems containing several 
$\Lambda$'s, $\Sigma$'s or $\Xi$'s. 
In fact, RMF explains spin-orbit interaction very nicely in 
normal nuclei as well as hypernuclei.
The contribution of spin-orbit interaction is very crucial in emerging the 
magic number in nucleonic sector and the same is expected in strange sector.

The paper is organized as follows: 
A brief introduction on hypernuclei and magic number is given in Section~\ref{int}. 
Section~\ref{form} gives a brief description of relativistic mean field formalism 
for hypernuclei with inclusion of $\Lambda N$ and $\Lambda \Lambda$ interactions.
The results are presented and discussed in Section~\ref{resu}.
The paper is summarized in Section~\ref{sum}.

\section{Formalism }\label{form}
Relativistic mean field theory has been applied succesfully 
to study the structural properties of normal nuclei as well as hypernuclei~\cite{rufa1990,glendenning1993,mares1994,sugahara1994,vretenar1998,lu2003,win2008,ikram14}.
The suitable expansion to hypernuclei has been made by including 
the lambda-baryon interaction Lagrangian with effective $\Lambda$N potential.
The total Lagrangian density for single-$\Lambda$ 
hypernuclei has been given in many Refs.~\cite{rufa1990,glendenning1993,mares1994,sugahara1994,vretenar1998,lu2003,win2008,ikram14}. 
For dealing the multi-lambda hypernuclei in quantitative way, the additional 
strange scalar ($\sigma^*$) and vector ($\phi$) mesons have been included 
which simulate the $\Lambda \Lambda$ interaction
~\cite{schaffner1994,schaffner1993,schaffner2002,shen2006}. 
Now, the total Lagrangian density can be written as
\begin{eqnarray} 
\mathcal{L}&=&\mathcal{L}_N+\mathcal{L}_\Lambda+\mathcal{L}_{\Lambda\Lambda} \;, 
\end{eqnarray}
\begin{eqnarray}
{\cal L}_N&=&\bar{\psi_{i}}\{i\gamma^{\mu}
\partial_{\mu}-M\}\psi_{i}
+{\frac12}(\partial^{\mu}\sigma\partial_{\mu}\sigma
-m_{\sigma}^{2}\sigma^{2})		   
-{\frac13}g_{2}\sigma^{3}                  \nonumber \\
&&-{\frac14}g_{3}\sigma^{4}
-g_{s}\bar{\psi_{i}}\psi_{i}\sigma 		
-{\frac14}\Omega^{\mu\nu}\Omega_{\mu\nu}
+{\frac12}m_{\omega}^{2}\omega^{\mu}\omega_{\mu}		\nonumber \\
&&-g_{\omega }\bar\psi_{i}\gamma^{\mu}\psi_{i}\omega_{\mu}    
-{\frac14}B^{\mu\nu}B_{\mu\nu} 
+{\frac12}m_{\rho}^{2}{\vec{\rho}^{\mu}}{\vec{\rho}_{\mu}}  
-{\frac14}F^{\mu\nu}F_{\mu\nu}                        \nonumber \\
&&-g_{\rho}\bar\psi_{i}\gamma^{\mu}\vec{\tau}\psi_{i}\vec{\rho^{\mu}} 
-e\bar\psi_{i}\gamma^{\mu}\frac{\left(1-\tau_{3i}\right)}{2}\psi_{i}A_{\mu}\;,  \nonumber \\
\mathcal{L}_{\Lambda}&=&\bar\psi_\Lambda\{i\gamma^\mu\partial_\mu
-m_\Lambda\}\psi_\Lambda
-g_{\sigma\Lambda}\bar\psi_\Lambda\psi_\Lambda\sigma  
-g_{\omega\Lambda}\bar\psi_\Lambda\gamma^{\mu}\psi_\Lambda \omega_\mu \;,  \nonumber \\
\mathcal{L}_{\Lambda\Lambda}&=&{\frac12}(\partial^{\mu}\sigma^*\partial_{\mu}\sigma^*  
-m_{\sigma^*}^{2}\sigma^{*{2}})
-{\frac14}S^{\mu\nu}S_{\mu\nu}
+{\frac12}m_{\phi}^{2}\phi^{\mu}\phi_{\mu}    \nonumber \\
&&-g_{\sigma^*\Lambda}\bar\psi_\Lambda\psi_\Lambda\sigma^*
-g_{\phi\Lambda}\bar\psi_\Lambda\gamma^\mu\psi_\Lambda\phi_\mu \;,      
\end{eqnarray}
where $\psi$ and $\psi_\Lambda$ denote the Dirac spinors for 
nucleon and $\Lambda$-hyperon, whose masses are M and
$m_\Lambda$, respectively.
Because of zero isospin, the $\Lambda$-hyperon does not couple 
to ${\rho}$- mesons.
The quantities $m_{\sigma}$, $m_{\omega}$, $m_{\rho}$, $m_{\sigma^*}$, 
$m_{\phi}$ are the masses of $\sigma$, $\omega$, $\rho$, $\sigma^*$, $\phi$ mesons 
and $g_s$, $g_{\omega}$, 
$g_{\rho}$, $g_{\sigma\Lambda}$, $g_{\omega\Lambda}$, $g_{\sigma^*\Lambda}$, 
$g_{\phi\Lambda}$ are their coupling constants, respectively. 
The nonlinear self-interaction coupling of ${\sigma}$ mesons is denoted 
by $g_2$ and $g_3$. 
The total energy of the system is given by 
$E_{total} = E_{part}(N,\Lambda)+E_{\sigma}+E_{\omega}+E_{\rho}
+E_{\sigma^*}+E_{\phi}+E_{c}+E_{pair}+E_{c.m.},$
where $E_{part}(N,\Lambda)$ is the sum of the single-particle energies of the 
nucleons (N) and hyperon ($\Lambda$).
The energies parts $E_{\sigma}$, $E_{\omega}$, $E_{\rho}$, $E_{\sigma^*}$, 
$E_{\phi}$, $E_{c}$, $E_{pair}$ and $E_{cm}$ are the contributions of meson 
fields, 
Coulomb field, pairing energy and the center-of-mass energy, respectively.
In present work, for meson-baryon coupling constant, NL3* parameter set is used 
through out the calculations~\cite{lalazissis09}. 
To find the numerical values of used $\Lambda-$meson coupling constants, 
we adopt the nucleon coupling to hyperon couplings ratio defined as; $R_\sigma=g_{\sigma\Lambda}/g_s$, $R_\omega=g_{\omega\Lambda}/g_\omega$, 
$R_{\sigma^*}=g_{\sigma^*\Lambda}/g_s$ and $R_\phi=g_{\phi\Lambda}/g_\omega$.
The relative coupling values are used as $R_\omega=2/3$, 
$R_\phi=-\sqrt{2}/3$, $R_\sigma=0.621$
and $R_{\sigma^*}=0.69$~\cite{chiapparini09,schaffner1994,dover1984}.
In present calculations, we use the constant gap BCS approximation to 
include the pairing interaction and the centre of mass 
correction is included by  $E_{cm}=-(3/4)41A^{-1/3}$.

\section{RESULTS AND DISCUSSIONS}\label{resu}
Before taking a detour on searching the $\Lambda$ magic behaviour in multi-$\Lambda$ 
hypernuclei, first we see the effects of introduced $\Lambda$ hyperon on normal 
nuclear core; how the binding energy and radii of 
normal nuclear system is affected by addition of $\Lambda$'s ?
To analyze this, we consider a list of normal nuclei covering a range from 
light to superheavy mass region i.e. $^{16}O$ to $^{378}120$.
Total binding energy (BE), binding energy per particle (BE/A), lambda binding 
energy ($B_\Lambda$) for $s-$ and $p-$ state and radii for considered 
core nuclei and corresponding hypernuclei are tabulated in Table~\ref{tab1}.
The calculated $B_\Lambda$ is compared with 
available experimental data and we found a close agreement between them.
This means the used parameter set is quite efficient to reproduce 
the experimental binding energy and ofcourse we can use it to make more 
calculations related to magicity in hypernuclei.
Since, we are dealing with closed shell hypernuclei and hence 
our RMF calculations is restricted to spherical symmetric.

The addition of $\Lambda$ hyperon 
to normal nuclei enhances the binding and shrinks the core of the system.
This is because of glue like role of $\Lambda$ hyperon that 
residing on the $s-$state for most of the time.
These observations are shown in Table~\ref{tab1}.
Binding energy of hypernuclei are larger than their normal 
counter parts and a reduction in total radius ($r_{total}$) of 
hypernuclei is observed, that means the $\Lambda$ particle 
makes the core compact with increasing binding. 
For example, the total radius of $^{16}O$ and $^{209}Pb$ is 
2.541 and 5.624 fm, which reduce to 2.536 and 5.616 fm by addition of 
single $\Lambda$ into the core of $O$ and $Pb$, respectively.
Moreover, for the sake of comparison with experimental data, binding 
energy and radii of the hypernuclei 
produced by replacing the neutrons means having a constant 
baryon number are also framed in Table~\ref{tab1} and the shrinkage effect is also noticed. 
This results show that an important impact of $\Lambda$ 
hyperon on binding as well as size of the system.
The increasing value of $B_\Lambda$ for 
$s-$state from light to superheavy hypernuclei confirming the potential 
depth of $\Lambda-$particle in nuclear matter which would be -28~MeV
~\cite{schaffner1994,millener1988}.

\begin{table*}
\caption{The calculated total binding energy and binding energy 
per particle for single-$\Lambda$ hypernuclei and their normal counter parts are listed here. 
The $\Lambda$ binding energy for $s-$ and $p-$state of considered 
hypernuclei are also mentioned and compared with available 
experimental values~\cite{hashimoto2006}, are given in parentheses. 
The radii are also displayed. Energies are given in unit of MeV and radii are in fm.}
\renewcommand{\tabcolsep}{0.20cm}
\renewcommand{\arraystretch}{1.3}
\begin{ruledtabular}
\begin{tabular}{cccccccccc}
Nuclei / Hypernuclei&BE &BE/A &$B_\Lambda^s$ &$B_\Lambda^p$ &$r_{ch}$ &$r_{total}$ &$r_p$ &$r_n$ &$r_\Lambda$ \\
\hline
$^{16}$O             &126.27 &7.89&      &      &2.674&         2.541&         2.555&         2.527&\\
$^{16}_\Lambda$O     &124.36 &7.77&-12.09 (12.5$\pm$0.35)&-2.66 (2.5$\pm$0.5)&2.673&         2.487&         2.550&         2.428&2.388\\
$^{17}_\Lambda$O     &137.99 &8.12&-11.98&-3.07 &2.673&         2.536&         2.554&         2.526&2.468\\
$^{40}$Ca            &341.43 &8.54&      &      &3.446&         3.331&         3.355&         3.307&\\
$^{40}_\Lambda$Ca    &344.16 &8.60&-17.51 (18.7$\pm$1.1)&-9.32 (11.0$\pm$0.6)&3.439&         3.292&         3.346&         3.262&2.693\\
$^{41}_\Lambda$Ca    &358.65 &8.75&-17.39&-9.46 &3.444&         3.315&         3.352&         3.304&2.737\\
$^{48}$Ca            &414.17 &8.63&      &      &3.444&         3.496&         3.359&         3.591&\\
$^{48}_\Lambda$Ca    &425.04 &8.86&-18.75&-10.95&3.439&         3.459&         3.353&         3.558&2.785\\
$^{49}_\Lambda$Ca    &433.01 &8.84&-18.94&-11.16&3.440&         3.479&         3.355&         3.586&2.791\\
$^{56}$Ni            &482.30 &8.61&      &      &3.696&         3.586&         3.610&         3.561&\\
$^{56}_\Lambda$Ni    &487.47 &8.71&-20.48&-12.67&3.695&         3.560&         3.609&         3.533&2.816\\
$^{57}_\Lambda$Ni    &502.96 &8.82&-20.72&-12.89&3.689&         3.567&         3.603&         3.555&2.817\\
$^{90}$Zr            &783.17 &8.70&      &      &4.249&         4.245&         4.179&         4.297&\\
$^{90}_\Lambda$Zr    &792.04 &8.80&-21.28&-15.25&4.246&         4.221&         4.175&         4.277&3.215\\
$^{91}_\Lambda$Zr    &804.69 &8.84&-21.37&-15.36&4.247&         4.233&         4.177&         4.295&3.222\\
$^{124}$Sn           &1048.19&8.45&      &      &4.642&         4.753&         4.580&         4.866&\\
$^{124}_\Lambda$Sn   &1060.81&8.55&-22.24&-17.10&4.633&         4.729&         4.571&         4.849&3.493\\
$^{125}_\Lambda$Sn   &1070.73&8.57&-22.28&-17.17&4.638&         4.742&         4.577&         4.864&3.503\\
$^{132}$Sn           &1102.69&8.35&      &      &4.689&         4.854&         4.631&         4.985&\\
$^{132}_\Lambda$Sn   &1118.13&8.47&-22.56&-17.65&4.680&         4.830&         4.620&         4.969&3.570\\
$^{133}_\Lambda$Sn   &1125.49&8.46&-22.60&-17.71&4.686&         4.843&         4.627&         4.983&3.579\\
$^{208}$Pb           &1639.32&7.88&      &      &5.499&         5.624&         5.448&         5.736&\\
$^{208}_\Lambda$Pb   &1655.85&7.78&-23.56 (26.3$\pm$08)&-19.74 (21.3$\pm$0.7)&5.492&         5.604&         5.441&         5.719&4.067\\
$^{209}_\Lambda$Pb   &1663.47&7.96&-23.54&-19.76&5.496&         5.616&         5.445&         5.734&4.076\\
$^{292}$120          &2063.09&7.06&      &      &6.271&         6.322&         6.225&         6.389&\\
$^{292}_\Lambda$120  &2103.33&7.12&-23.73&-20.94&6.262&         6.306&         6.216&         6.376&3.316\\
$^{293}_\Lambda$120  &2111.67&7.21&-23.59&-20.86&6.268&         6.316&         6.223&         6.387&3.320\\
$^{304}$120          &2140.81&7.04&      &      &6.302&         6.417&         6.258&         6.519&\\
$^{304}_\Lambda$120  &2184.09&7.19&-23.89&-21.03&6.298&         6.403&         6.253&         6.507&3.272\\
$^{305}_\Lambda$120  &2189.18&7.18&-23.79&-20.96&6.300&         6.411&         6.255&         6.518&3.301\\
$^{378}$120          &2385.44&6.31&      &      &6.714&         7.144&         6.678&         7.350&\\
$^{378}_\Lambda$120  &2432.02&6.11&-23.09&-20.81&6.896&         7.344&         6.863&         7.543&3.642\\
$^{379}_\Lambda$120  &2433.89&6.09&-23.43&-21.04&6.712&         7.138&         6.676&         7.350&3.549\\
\end{tabular}
\end{ruledtabular}
\label{tab1}
\end{table*}

\begin{figure}
\includegraphics[width=1.0\columnwidth]{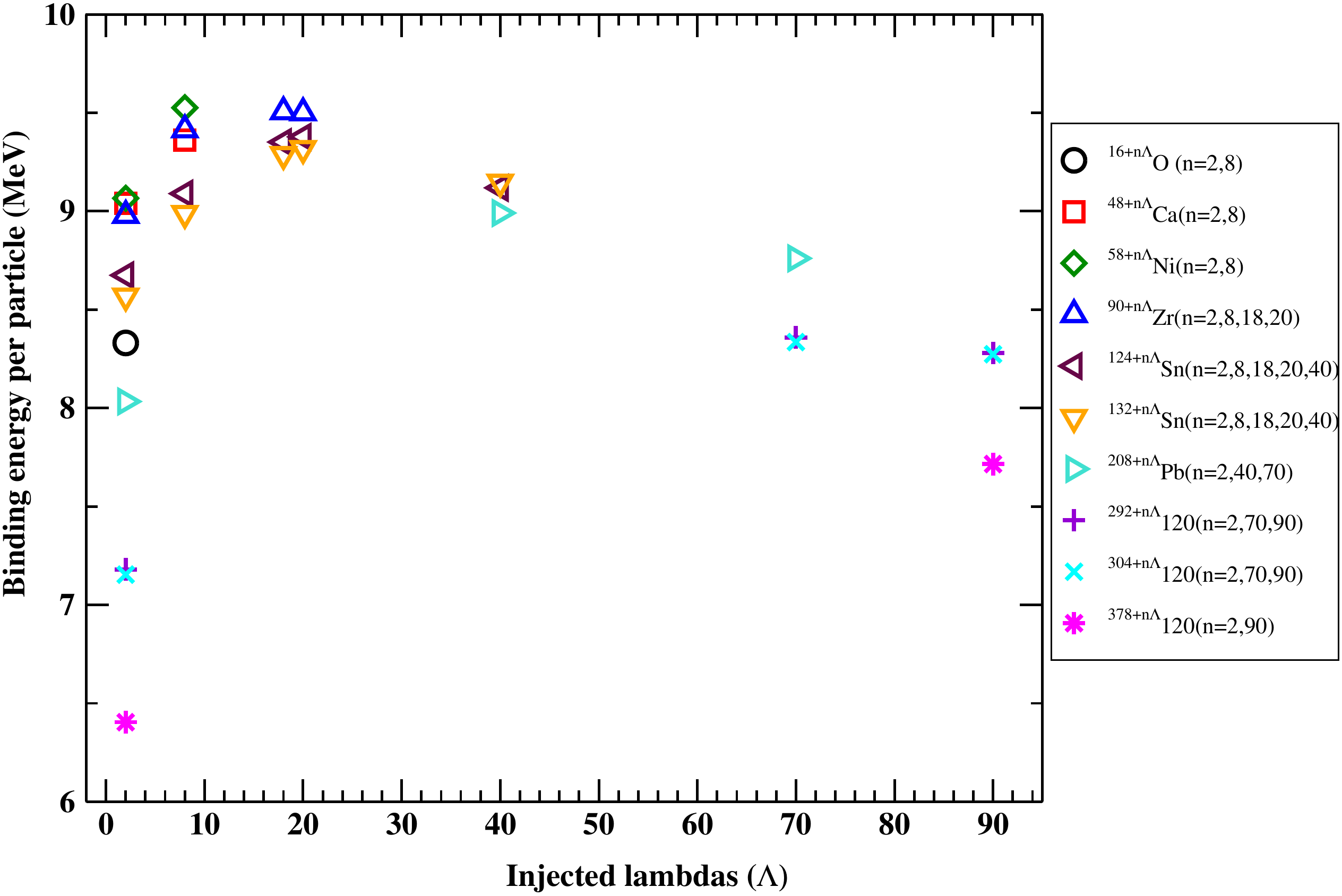}
\caption{\label{be_perb}(color online) Binding energy per particle for triply 
magic multi-$\Lambda$ hypernuclei.}
\end{figure}

\subsection{Stability of hypernuclei} 
Binding energy provides the detailed information of various elements corresponding 
to their stability. 
Binding energy per particle increases upto the element iron whose atomic 
number is 26 and mass number 57. 
The information provided by the binding energy per particle curve is that 
iron and its neighbouring elements (Ni) are most stable i.e. they neither 
undergo fission or fusion. 
Thus the significance of the binding energy per particle curve lies in the 
fact that it is an indicator of nuclear stability and thus helps in 
classifying the elements which undergo fission, fusion and radioactive 
disintegration.  
We noticed a similar pattern of binding energy per particle in hypernuclear 
regime also. 
The triply magic system is produced by addition of $\Lambda$ magic 
number into the core of doubly shell closure such as 
$^{16}O$,$^{48}Ca$,$^{58}Ni$,$^{90}Zr$,$^{124}Sn$,$^{132}Sn$,$^{208}Pb$,
$^{292}120$,$^{304}120$, $^{378}120$.
Binding energy per particle of considered systems confirming that nickel 
with 8 $\Lambda$'s ($^{56+8\Lambda}Ni$)
being the most tightly bound hypernucleus (BE/A = 9.5 MeV) as shown in Fig.~\ref{be_perb}.
This results are in remarkable agreement with earlier predictions~\cite{schaffner1994, schaffner1993}.

\begin{figure}
\includegraphics[width=1.0\columnwidth]{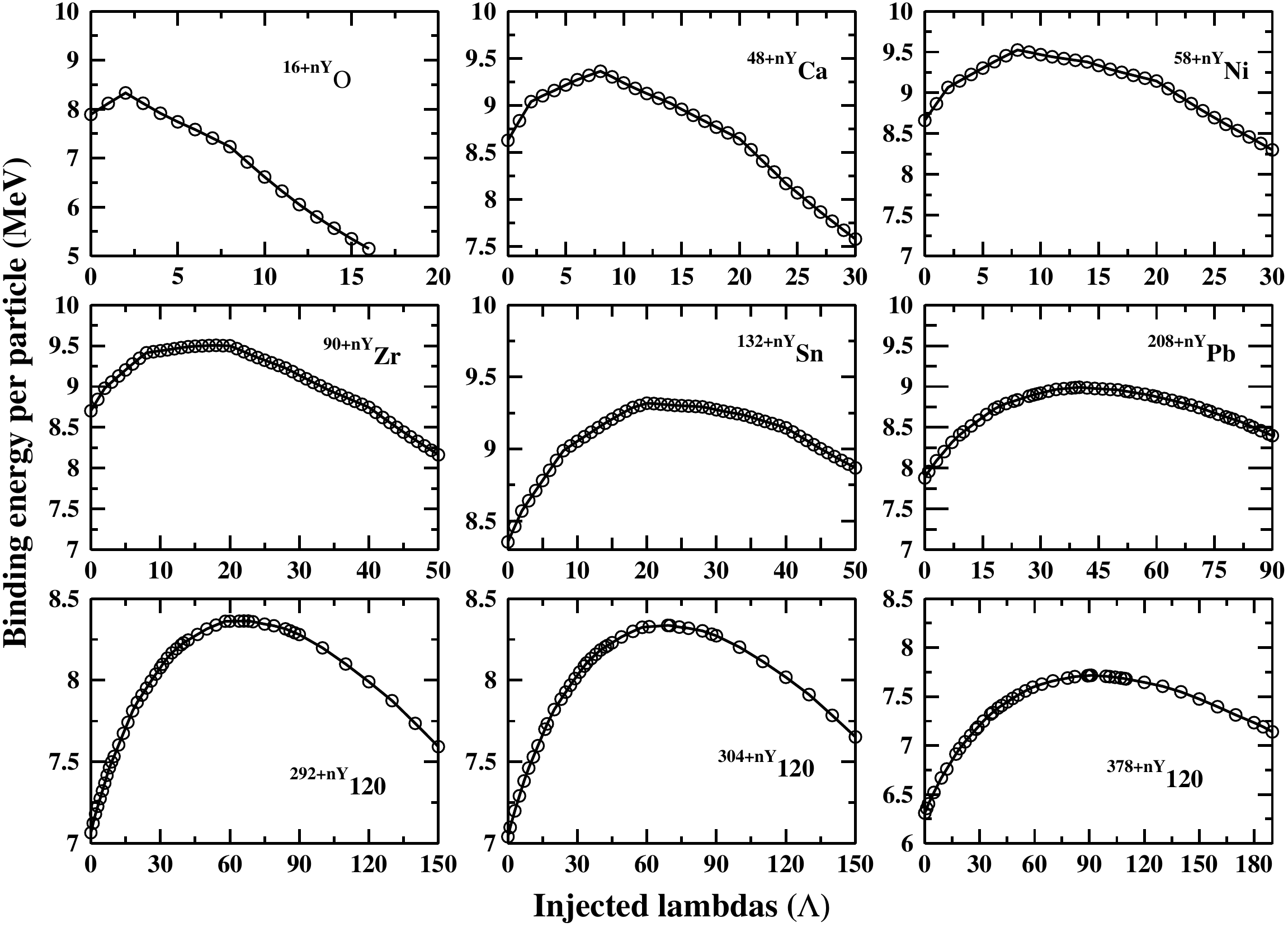}
\caption{\label{total_be}Binding energy per particle for light 
to supeheavy mass multi-$\Lambda$ hypernuclei.}
\end{figure}
 
\subsection{Binding energy and separation energy} 
\begin{figure}
\includegraphics[width=1.0\columnwidth]{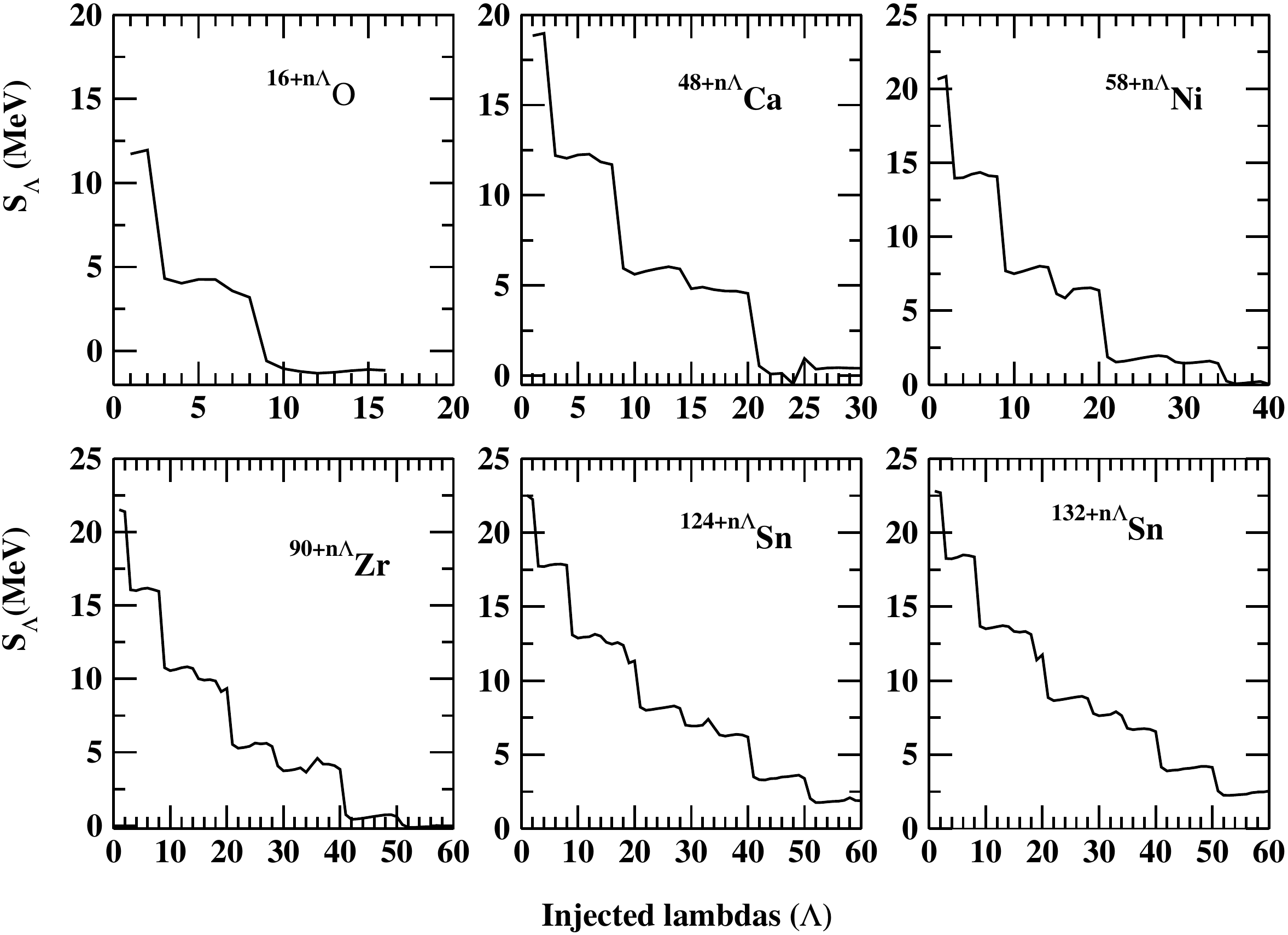}
\caption{\label{s1-light}One lambda separation energy for 
medium mass multi-$\Lambda$ hypernuclei.}
\end{figure}

\begin{figure}
\includegraphics[width=1.0\columnwidth]{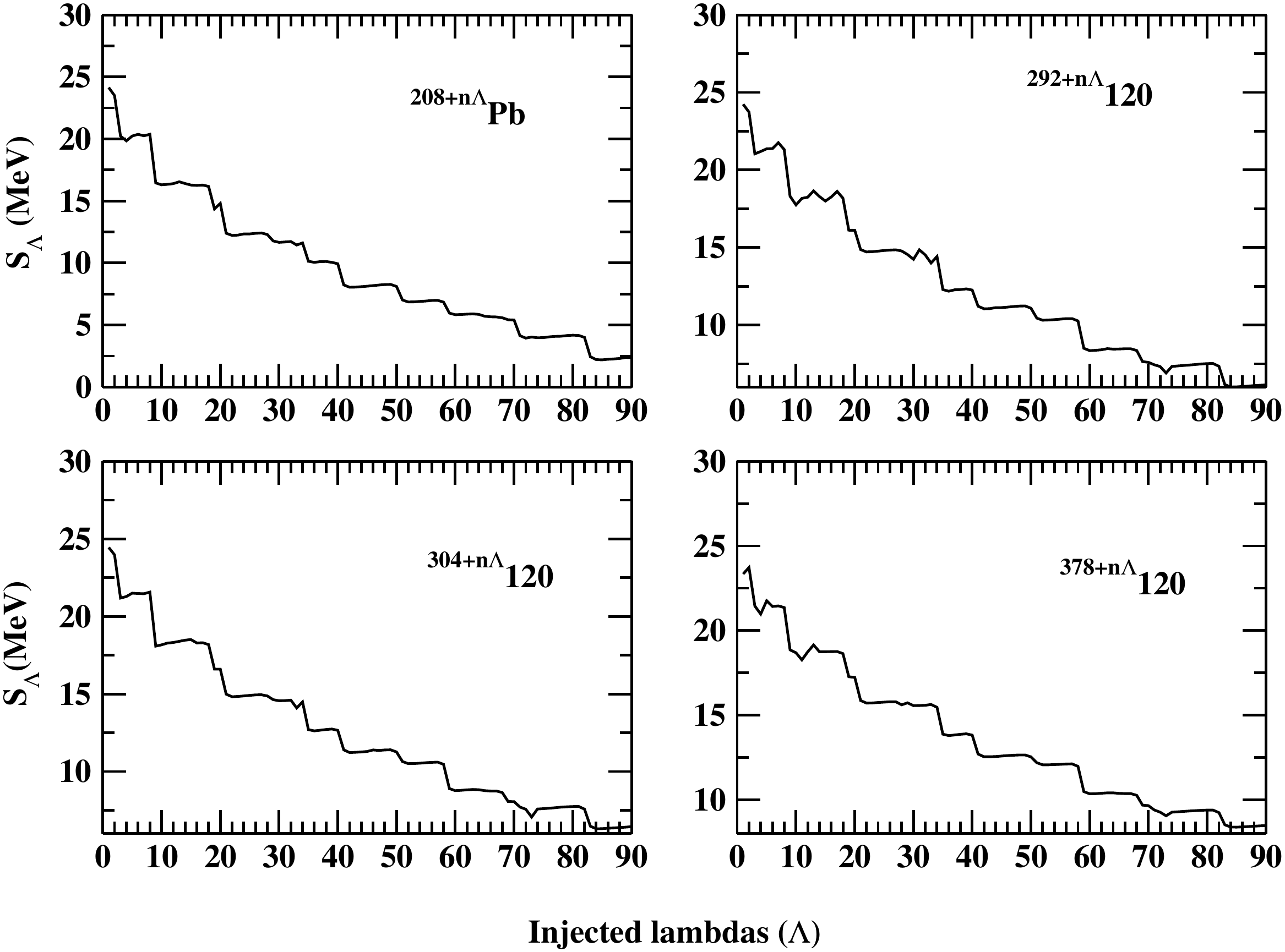}
\caption{\label{s1-heavy}Same as Fig.~\ref{s1-light} but for heavy 
to superheavy mass multi-$\Lambda$ hypernuclei.}
\end{figure}

\begin{figure}
\includegraphics[width=1.0\columnwidth]{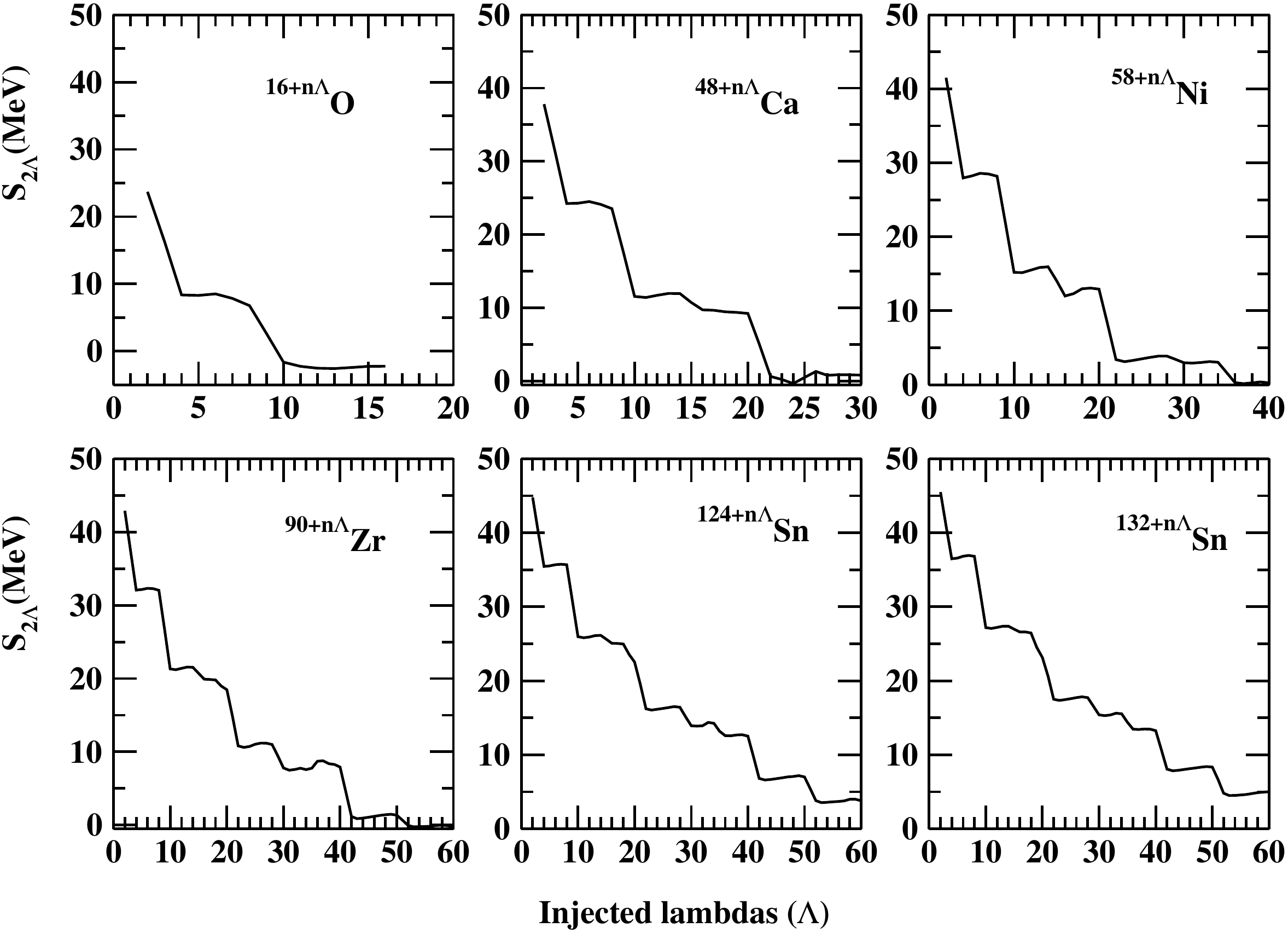}
\caption{\label{s2-light}Two lambda separation energy for 
medium mass multi-$\Lambda$ hypernuclei.}
\end{figure}

\begin{figure}
\includegraphics[width=1.0\columnwidth]{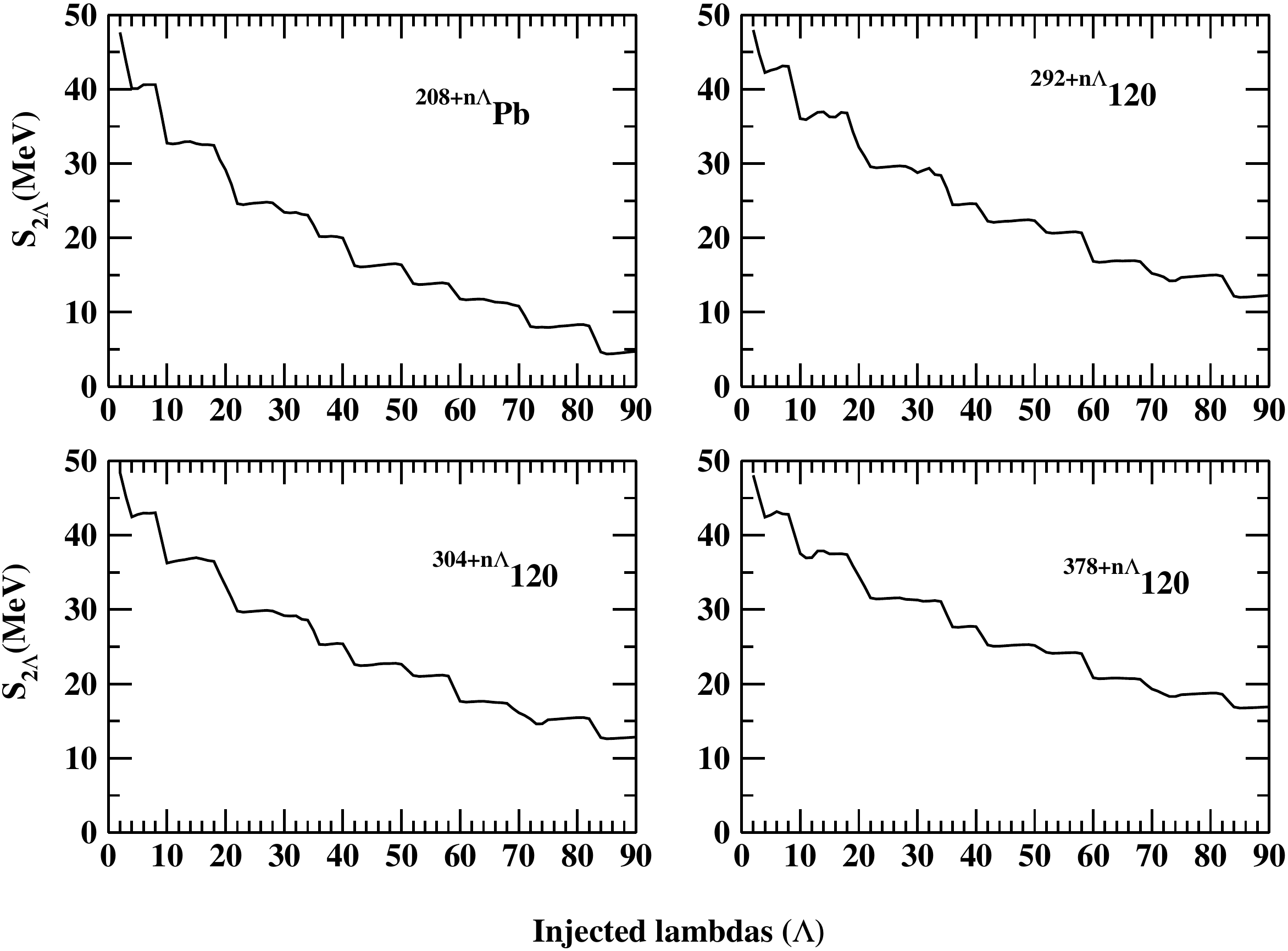}
\caption{\label{s2-heavy}Same as Fig.~\ref{s2-light} but for heavy 
to superheavy mass multi-$\Lambda$ hypernuclei. }
\end{figure}

To analyze the magic behaviour of lambda in multi-$\Lambda$ hypernuclei, we choose the nuclear core of 
doubly shell closure including predicted shell closure nuclei of superheavy mass region 
and than added the $\Lambda$ hyperons. 
Initially, we look for the binding energy per particle with respect to added 
hyperons for considered hypernuclei and plotted in Fig.~\ref{total_be}.
The peak value of the graph corresponds to maximum stability for a particular 
hypernuclear system. 
It also explains that the injection of few $\Lambda$ hyperons 
enhance the binding of the light mass hypernuclei, conversely 
further addition reduces the binding. 
However for heavy mass region, the BE/A increases with addition of 
large number of hyperons and form a most bound system but 
further addition decreases the binding energy gradually.
This means certain number of added $\Lambda$'s to a particular nuclear core 
form a most stable system.
For example, the injection of 2 $\Lambda$'s provide a maximum stability to 
$^{16}O$.
Along the similar line a maximum binding is observed 
for $^{48}Ca$ with $\Lambda=$ 8 and this number goes to 90 for superheavy core.
In this way, we extract certain numbers of added $\Lambda$'s that 
is 2, 8, 18, 20, 40, 70, 90 which provide maximum stability for the considered systems 
(i.e. $^{16+2\Lambda}O$, $^{48+8\Lambda}Ca$, $^{58+8\Lambda}Ni$, 
$^{90+18\Lambda}Zr$, $^{124+20\Lambda}Sn$, $^{132+20\Lambda}Sn$, 
$^{208+40\Lambda}Pb$, $^{292+68\Lambda}120$, $^{304+70\Lambda}120$, $^{378+90\Lambda}120$) 
and these numbers may correspond to $\Lambda$ magic number in multi-$\Lambda$ hypernuclei.
But, there exist several other strong signatures of marking the magic 
number such as; separation energy, shell gaps, pairing energy etc.
Therefore, to analyze the actual behaviour of magicity, we make 
the analysis of such relevant parameters.
In this regard, we estimate one- and two-lambda separation energy 
$S_\Lambda$, $S_{2\Lambda}$, which are known to be first insight of shell closure.
In analogy of nucleonic sector, the magic number in multi-hypernuclei 
may characterized by the large lambda shell gaps in single-particle energy levels.
The extra stability given by certain number of introduced $\Lambda$'s can also 
be detect from sudden fall in $\Lambda$ separation energy.
Therefore, on quest of magicity in multi-$\Lambda$ hypernuclei, one- and two-lambda 
separation energy is estimated using the following expressions;
\begin{equation*}
S_\Lambda(N,Z,\Lambda)=BE(N,Z,\Lambda)-BE(N,Z,\Lambda-1)
\end{equation*}
and 
\begin{equation*}
S_{2\Lambda}(N,Z,\Lambda)=BE(N,Z,\Lambda)-BE(N,Z,\Lambda-2).
\end{equation*}
These quantities are plotted in Fig.~\ref{s1-light},~\ref{s1-heavy},~\ref{s2-light} and~\ref{s2-heavy}.
For a lambda chain, the $S_\Lambda$ and $S_{2\Lambda}$ becomes larger with increasing 
number of lambda $\Lambda$.
For a fixed Z, N; $S_\Lambda$ and $S_{2\Lambda}$ decrease gradually with lambda number.
A sudden decrease of $S_\Lambda$ and $S_{2\Lambda}$ just after the magic number in lambda 
chain like as neutron chain indicates the occurrence of $\Lambda$ shell closure.
The sudden fall of $S_\Lambda$ at 
$\Lambda=$ 2, 8, 14, 18, 20, 28, 34, 40, 50, 58, 68, 70 
and 82 can clearly be seen in considered hypernuclear candidates revealing 
a signature of magic character.
Moreover, $\Lambda=$ 14 and 28 are observed only in light mass mulit-$\Lambda$ 
hypernuclei, even $\Lambda=$ 28 does not show pronounced energy separation.
However, a good strength of sudden fall of $S_\Lambda$ at $\Lambda=$ 34 
and 58 is clearly observed in heavy and superheavy mass region.

Two-lambda separation energy provides more strong signature to quantify shell 
closure due to absence of odd-even effects. 
Figures~\ref{s2-light} and \ref{s2-heavy} reveal that sudden fall of $S_{2\Lambda}$ 
at $\Lambda=$ 2, 8, 14, 18, 20, 28, 34, 40, 50, 58, 68, 70 and 82 as 
observed in considered multi-$\Lambda$ hypernuclear candidates. 
These certain numbers corresponds to $\Lambda$ magic number in multi-$\Lambda$ hypernuclei 
and form a triply magic system with doubly magic core.
This is the central theme of the paper.
The significant falls of $S_\Lambda$ and $S_{2\Lambda}$ at $\Lambda=$14 is 
appeared in Ca and Ni hypernuclei. 
The lambda number 28 seems to be very much feeble magic number, contrary to nucleonic sector.
The another new lambda number 68 suppose to be semi-magic arises due to 
subshells closure.
For the sake of clear presentation of the results, we also make the analysis for 
two-lambda shell gaps ($\delta_{2\Lambda}$) and plotted as a function of added $\Lambda$'s.

Summarizing the above results, we may say that based on one- and two-lambda separation 
energies $S_\Lambda$ and $S_{2\Lambda}$, the signatures of the magicity in RMF 
appears at 2, 8, 14, 18, 20, 28, 34, 40, 50, 58, 68, 70 and 82.
The lambda number 28 and 68 are appeared in light and heavy hypernuclei, respectively 
and suppose to be feeble magic number.

\subsection{Two-lambda shell gap}
\begin{figure}
\includegraphics[width=1.0\columnwidth]{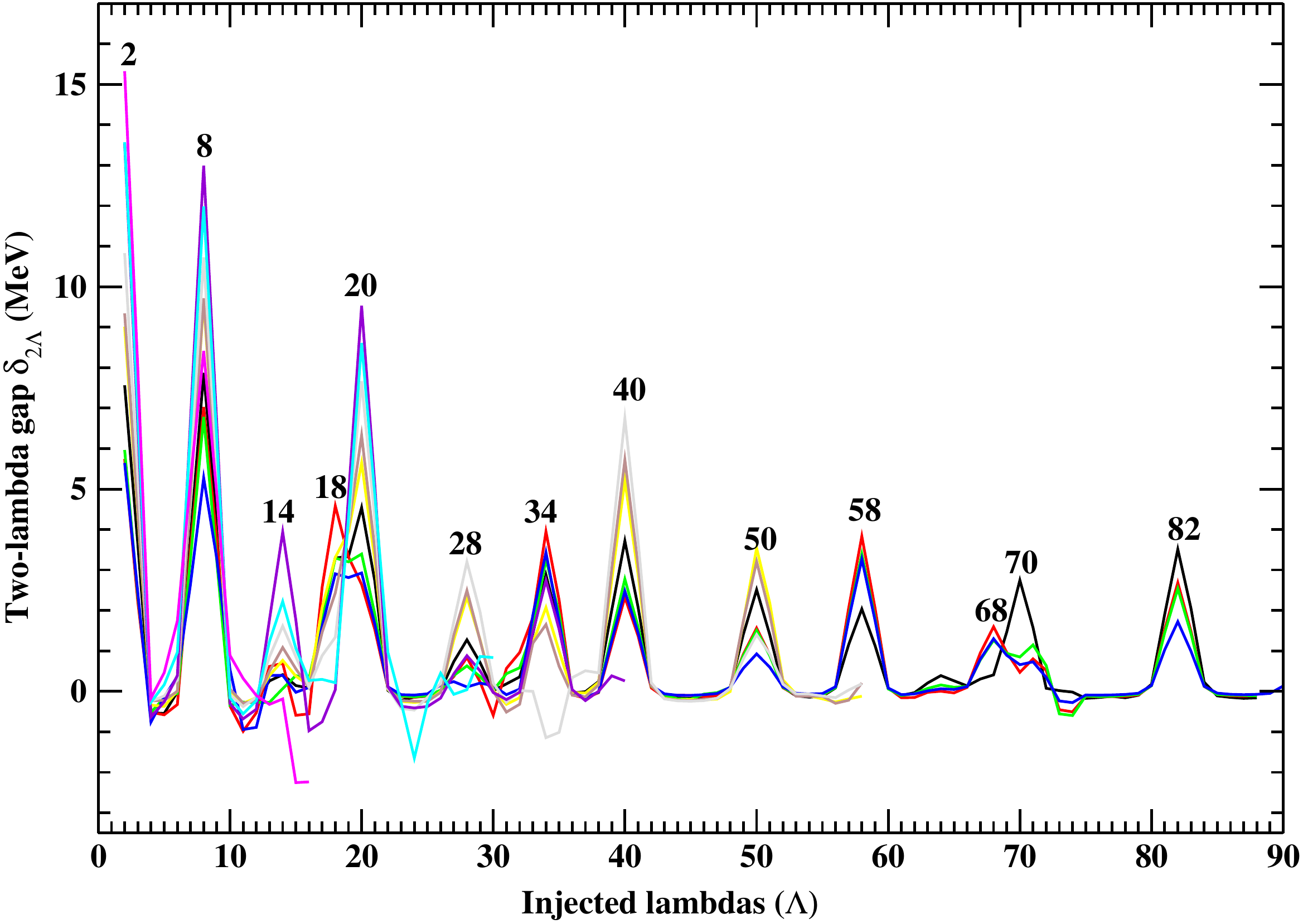}
\caption{\label{s-gaps}(color online) Two lambda shell gap is shown for 
considered multi-$\Lambda$ hypernuclei. }
\end{figure}

The change of the two-lambda separation energies can also be quantified 
by the second difference of the binding energies, i.e., two-lambda shall gap 
which is expressed by:
\begin{small}
\begin{eqnarray*}
\delta_{2\Lambda}(N,Z,\Lambda)&=&2BE(N,Z,\Lambda)-BE(N,Z,\Lambda+2)-BE(N,Z,\Lambda-2) \\
			      &=&S_{2\Lambda}(N,Z,\Lambda)-S_{2\Lambda}(N,Z,\Lambda+2).			
\end{eqnarray*}
\end{small}
A peak of two-lambda shell gaps indicates the drastic change of the two-lambda 
separation energies; which is used as one of the significant signature of magic number.
The two-lambda shell gaps $\delta_{2\Lambda}$, for all considered hypernuclei as a 
function of added $\Lambda$ hyperons are shown in Fig~\ref{s-gaps}.
A peak at certain $\Lambda$ number suggests the existence of lambda shell closure. 
However, the quality of magic number is represented  by sharpness as 
well as magnitude of the peak. 
Figure.~\ref{s-gaps} reveals that the magnitude of the peak is found 
to be largest at $\Lambda=$ 2, 8, 20, 40 indicating the strong shell closures.
Further, the peaks appeared at $\Lambda=$ 14, 18, 28, 34, 50, 58, 70 and 82 
indicate the respective lambda magic number.
Moreover, a peak with a very small magnitude is also appeared at $\Lambda=$ 68 due 
to closureness of subshell (2$d_{3/2}$) revealing $\Lambda$ semi-magic number.
A peak with small magnitude is seen at $\Lambda$= 28 
representing a feeble lambda magic number, contrary to nucleonic magic number.
Pronounced peak is appeared at $\Lambda=$ 34 and 58 indicating a strong $\Lambda$ closed shells.

\subsection{Density profile and bubble structure} 
\begin{figure}
\includegraphics[width=1.0\columnwidth]{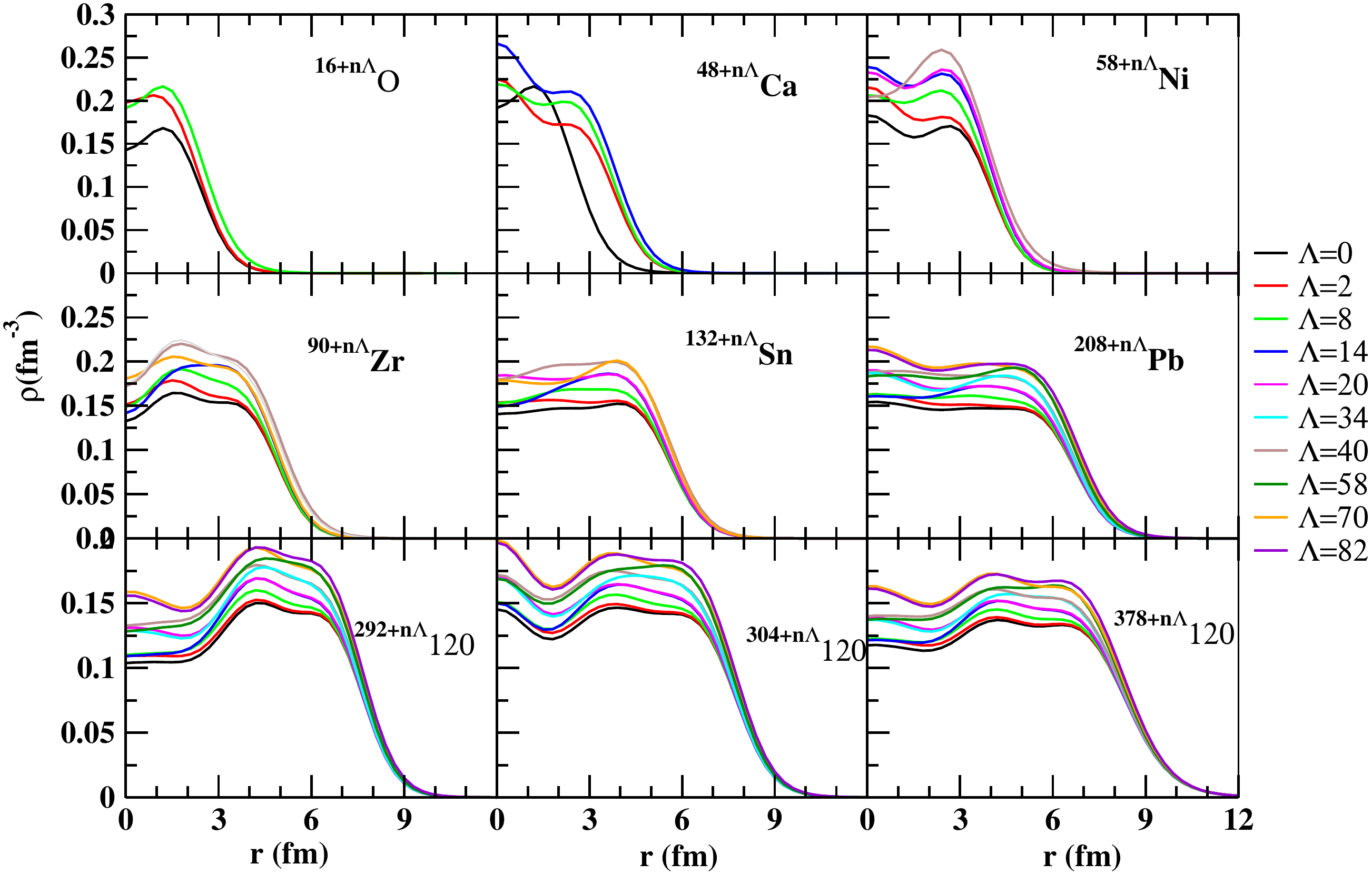}
\caption{\label{density}(color online) Total (Nucleon plus lambda) density 
for considered triply magic hypernuclei with $\Lambda$=0, 2, 8, 14, 20, 34, 40, 58, 70 and 82.}
\end{figure}

A hypernucleus is a composed system of nucleons and hyperons and hence the 
gross structure of hypernucleus can be described by density distribution 
of nucleons as well as hyperons.
It is well known and has mentioned earlier that the addition of a $\Lambda$ hyperon  
makes the nuclear core compact with increasing binding as well as density.
Therefore, it is important to study the effects of large number of added 
$\Lambda$ hyperons on the nuclear density.
Due to addition of hyperons, the magnitude of total density increases with increasing 
number of $\Lambda$'s as shown in Fig.~\ref{density}. 
On view the density profile, one can examine the most interesting 
feature of nuclei i.e. bubble structure, which measure the depression 
of central density and has already been observed in light to 
superheavy mass region~\cite{grasso2009,ekhan08,decharge03,ikram14}.
It is to be noticed that several factors, including pairing correlations
~\cite{grasso2009}, tensor force~\cite{wang2013,nakada2013} and 
dynamic shape fluctuations~\cite{yao2012,yao2013,wu2014} turn out to have 
influence on depression of central density.
The exotic structures like bubble and halo have been 
recently studied in $\Lambda-$hypernuclei~\cite{ikram14,ekhan2015}. 

Owing to weaker $\Lambda \Lambda$ attraction compared to the nucleon-nucleon one the 
lambda hyperons are more diffuse in a nucleus than nucleons and thus generating a hyperon 
density about $1/3$ smaller than the nucleonic density. 
Thus, it becomes quite important to look for the effect of large number of hyperons on neutron and 
proton density distributions. 
Since, there is no change of nucleon number and hence no anomalous effect 
of introduced $\Lambda$'s on neutron, proton densities is observed, individually. 
But the total density of the system is largely affected due to 
increasing number of $\Lambda$'s into the core. 
In considered multi-hypernuclei, the nucleonic core of some of them 
shows the depression of central density for example 
$^{16}O$, $^{90}Zr$, $^{292}120$, $^{304}120$ and $^{378}120$ as predicted 
earlier also~\cite{decharge03,ikram14}.
It is found that the injected $\Lambda$'s reduce the 
depression of central density. 
For example, the depression of central part in $^{16}O$ is reduced by 
injection of 2 $\Lambda$'s and further more by 8 $\Lambda$'s.
The $\Lambda$ particle attracts the nucleons 
towards the centre enhancing the central density and as a 
result remove the bubble structure partially or fully 
as reflected in Fig.~\ref{density}.
Therefore, it is one of the important implication of 
$\Lambda$ particle to the nuclear system.
Beyond the bubble structure, no anomalous behaviour of total 
density (core + $\Lambda$) in triply magic system is reported.
 
\subsection{Spin-orbit interaction and mean field potentials} 
\begin{figure}
\includegraphics[width=1\columnwidth,height=8cm]{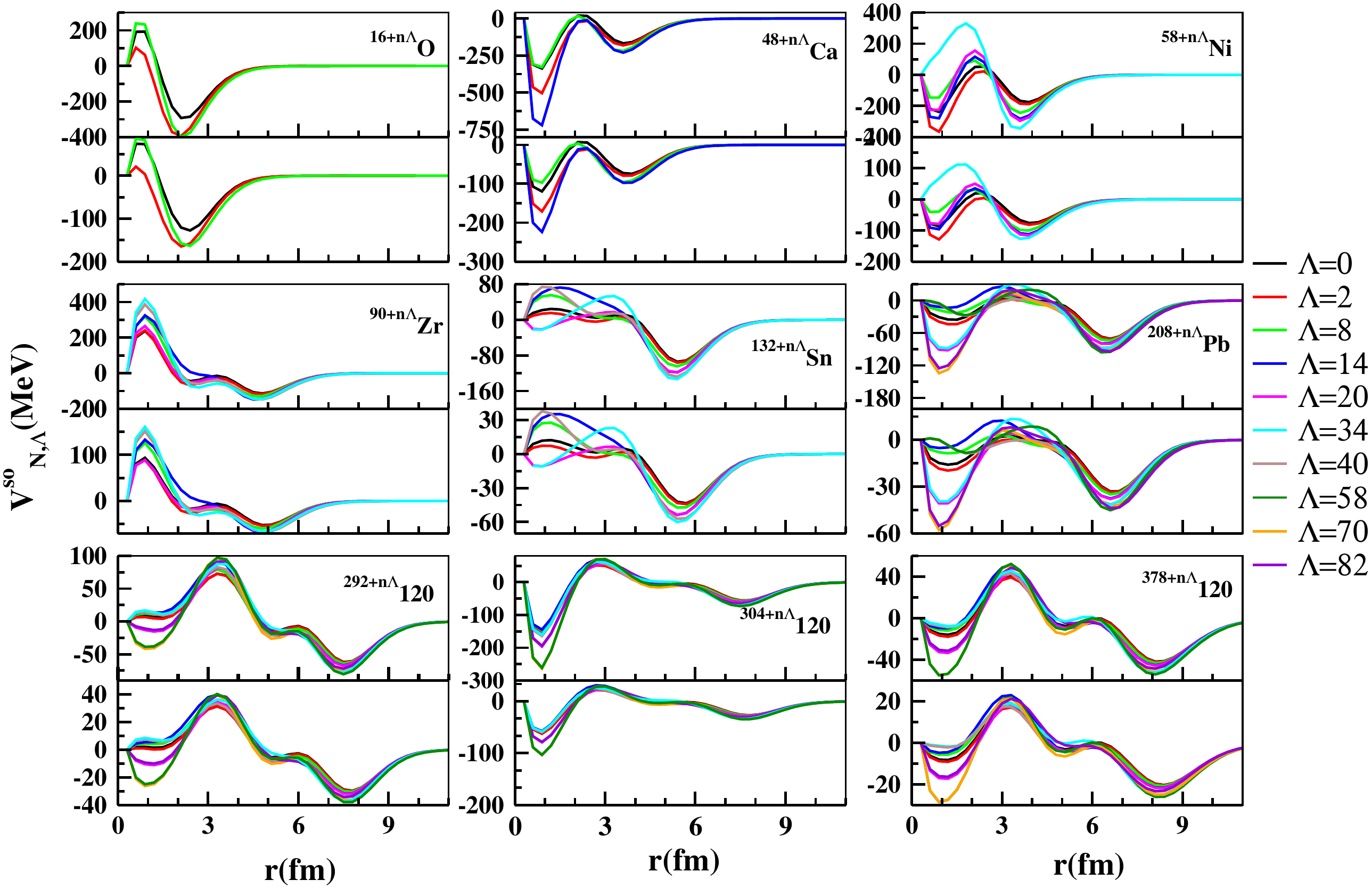}
\caption{\label{potentials}(color online) Spin-orbit interaction potentials 
of nucleon and lambda for considered triply magic hypernuclei 
with $\Lambda$=0, 2, 8, 14, 20, 34, 40, 58, 70 and 82.
The upper part in each panel represents the nucleonic spin-orbit and 
the lower one representing the lambda spin-orbit interaction.}
\end{figure}

The spin-orbit interaction plays a significant role in reproducing 
the results quantitatively.  
It is the beauty of RMF in which the spin-orbit splitting 
is built-in naturally with exchange of scalar and vector mesons 
and thus describe a nuclear fine structure.
It is not limited only to nuclei or superheavy nuclei but appears  
in hypernuclei also, however the strength of interaction is weaker 
than normal nuclei~\cite{boguta1981,noble1980,vretenar1998}.
It is clearly seen from Figs.~\ref{potentials} that 
the spin-orbit potential for lambda hyperon is weaker than their 
normal counter parts and our results are consistent with 
theoretical predictions and experimental 
measurements~\cite{brockmann1977,keil2000,ajimura2001}.
Here, nucleon ($V^N_{so}$) and lambda ($V^\Lambda_{so}$) 
spin-orbit interaction potentials are calculated for 
considered triply magic multi-hypernuclei.
It is also conclude that the addition of $\Lambda$'s 
affects the nucleon as well as $\Lambda$ spin-orbit potential to a great extent.

\begin{figure}
\includegraphics[width=1.0\columnwidth]{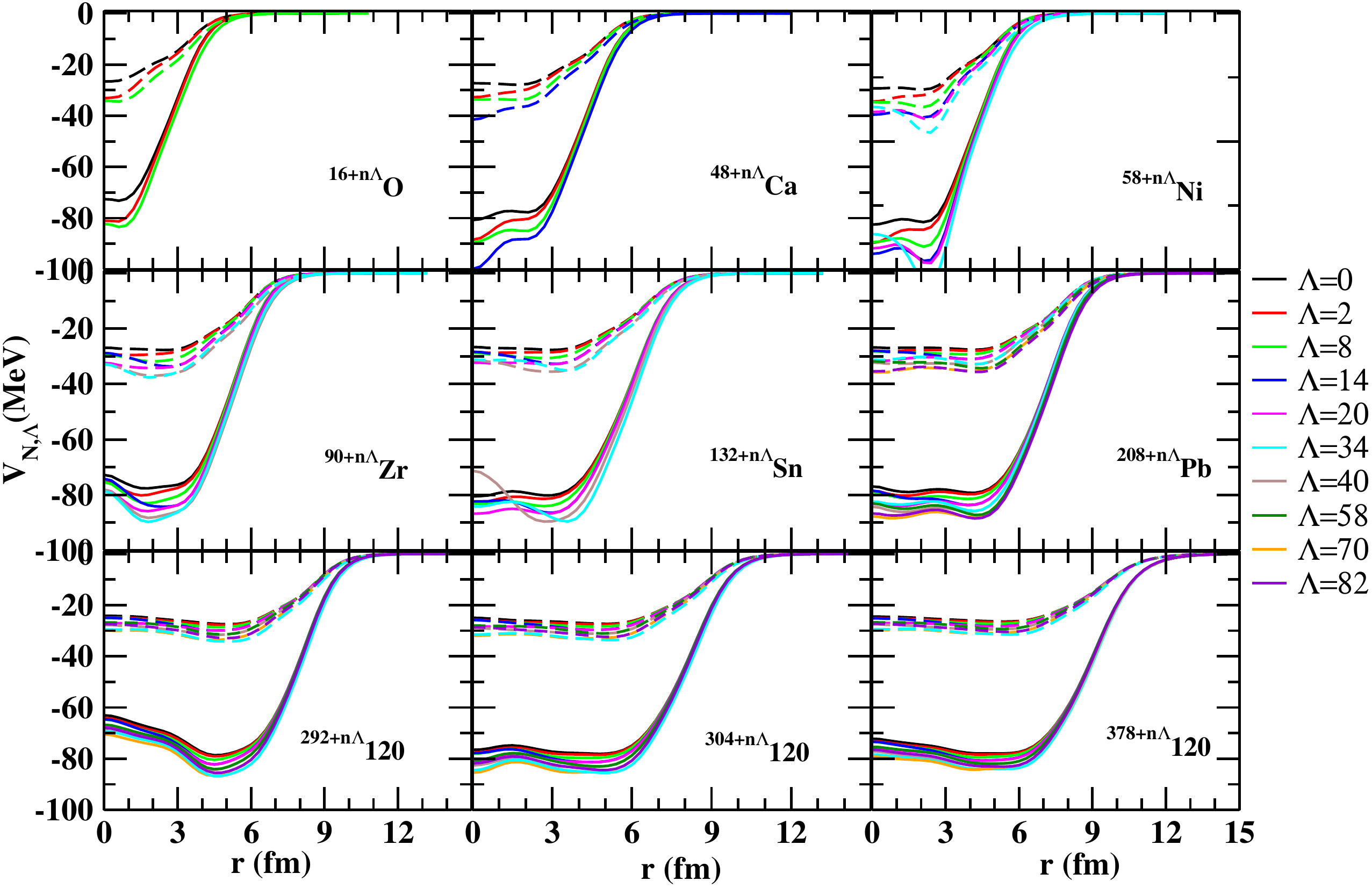}
\caption{\label{meanpot}(color online) Total mean field potentials for 
considered triply magic hypernuclei with 
$\Lambda$=0, 2, 8, 14, 20, 34, 40, 58, 70 and 82.
Dashed lines represents the lambda mean potential and the 
nucleon potential is shown by solid line.}
\end{figure}

The nucleon ($V_N$) and lambda ($V_\Lambda$) mean field potentials are also 
investigated and plotted as a function of radial parameter 
shown in Fig.~\ref{meanpot}.
The total depth of $\Lambda$ mean field potential is found to be around -30 MeV, 
which is in agreement with existing experimental data~\cite{keil2000}.
It is to mention that the additions of $\Lambda$'s affects the depth of 
both nucleon as well as lambda mean potentials.
The nucleonic potential depth in multi-lambda hypernuclei is approx 
to -80 MeV to -90 MeV.
The shape of lambda potential looks like to be same as 
nucleonic potential and only the amount of depth is different.
It is also to be noticed that the nucleonic potential looks like as 
V-shape type and shows the maximum depth around -90 MeV at r = 4 fm, 
while this amount of depth reaches to -70 MeV at r = 0 fm for $^{292}$120.
It indicates a relatively low concentration 
of the particles at central region (r = 0) which is a direct 
consequence of depression of central density so-called bubble structure.

\subsection{Single-particle energies} 
\begin{figure}
\includegraphics[width=1.0\columnwidth]{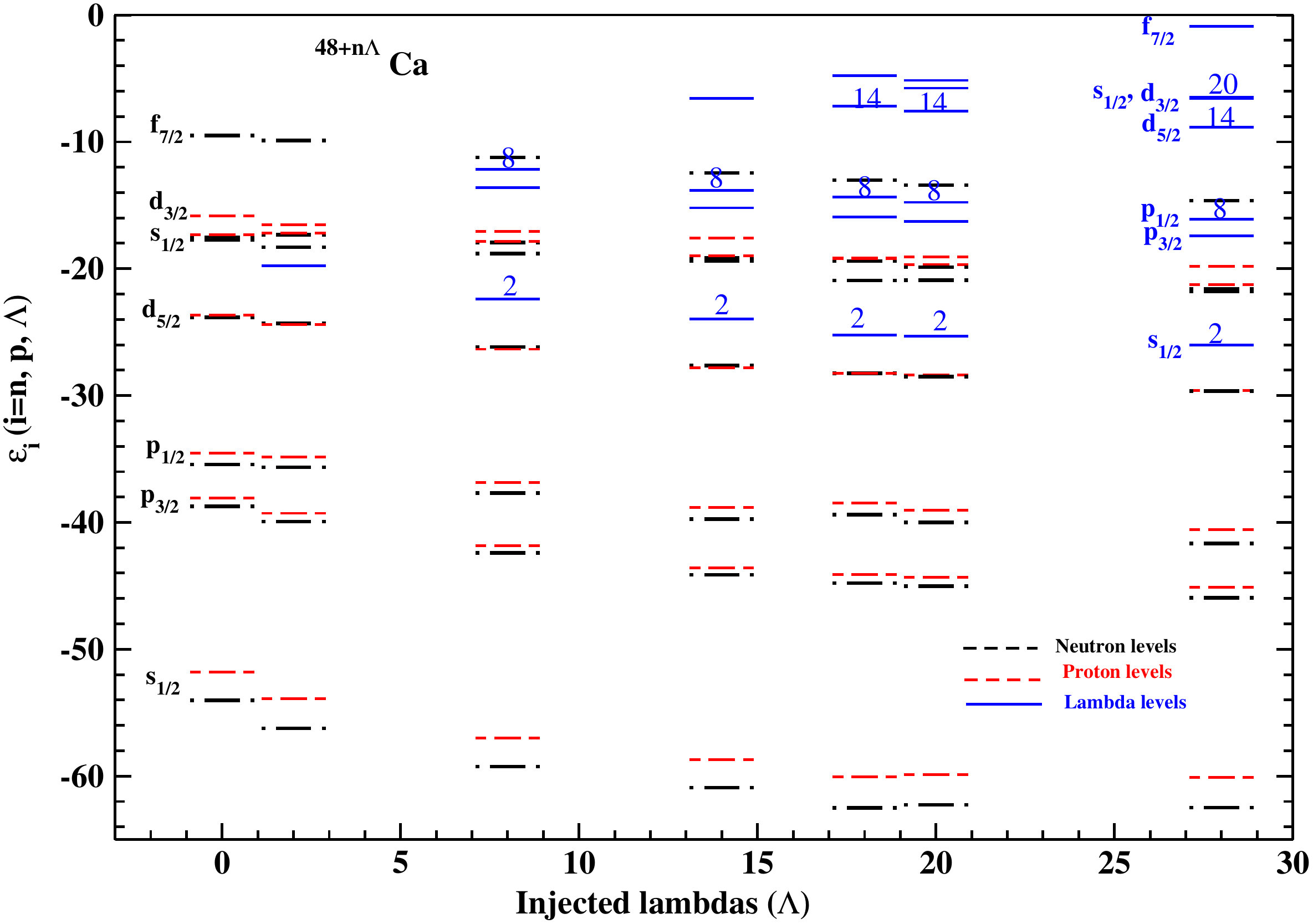}
\caption{\label{levels-ca}(color online) Single-particle energy levels for 
triply magic Ca multi-hypernuclei for $\Lambda=$ 2, 8, 14, 18, 20 and 28.}
\end{figure}

Any kinds of change in a system can be observed from their 
single-particle energy levels.  
To analyze the impact of $\Lambda$ hyperon on nucleon single-particle energy levels, 
the filled neutron and proton levels for Ca hypernuclei are 
plotted as a function of added hyperons as shown in Fig.~\ref{levels-ca}.
Figure~\ref{levels-ca} reveals that the neutron and proton energy levels 
goes dipper with addition of $\Lambda$'s as a result increase the stability of the system.
The added hyperons increase the nucleon separation energy  and as a result form a 
more bound system with increasing binding then their normal counter parts, 
which also leads to an extension of drip-line~\cite{Samanta2008}.
For example, neutron $s_{1/2}$(n) level has an energy of about -54.037 MeV for
the core of Ca hypernucleus, while this amount reaches to -62.499 MeV for 
$^{48+18\Lambda}Ca$ system with 18 $\Lambda$'s. 
Also, a same trend is observed for proton 
levels where, $s_{1/2}$(p) has an energy 51.787 MeV for the core 
of Ca hypernuclei and this value reaches to -60.0477 MeV with 
addition of 18 $\Lambda$'s.
This results show that the $\Lambda$ hyperons draw the nuclear system towards 
more stability with increasing strangeness.
Moreover, the same trend of neutron and proton energy levels is observed for other 
multi hypernuclei where both the levels would go dipper with increasing 
number of $\Lambda$ hyperon to nucleonic core but we do not make a plot for the same.
A inversion of proton levels is seen, where $d_{3/2}$ fill faster than $s_{1/2}$ and 
this type of filling is also observed in lambda levels.

\begin{figure}
\includegraphics[width=1.0\columnwidth]{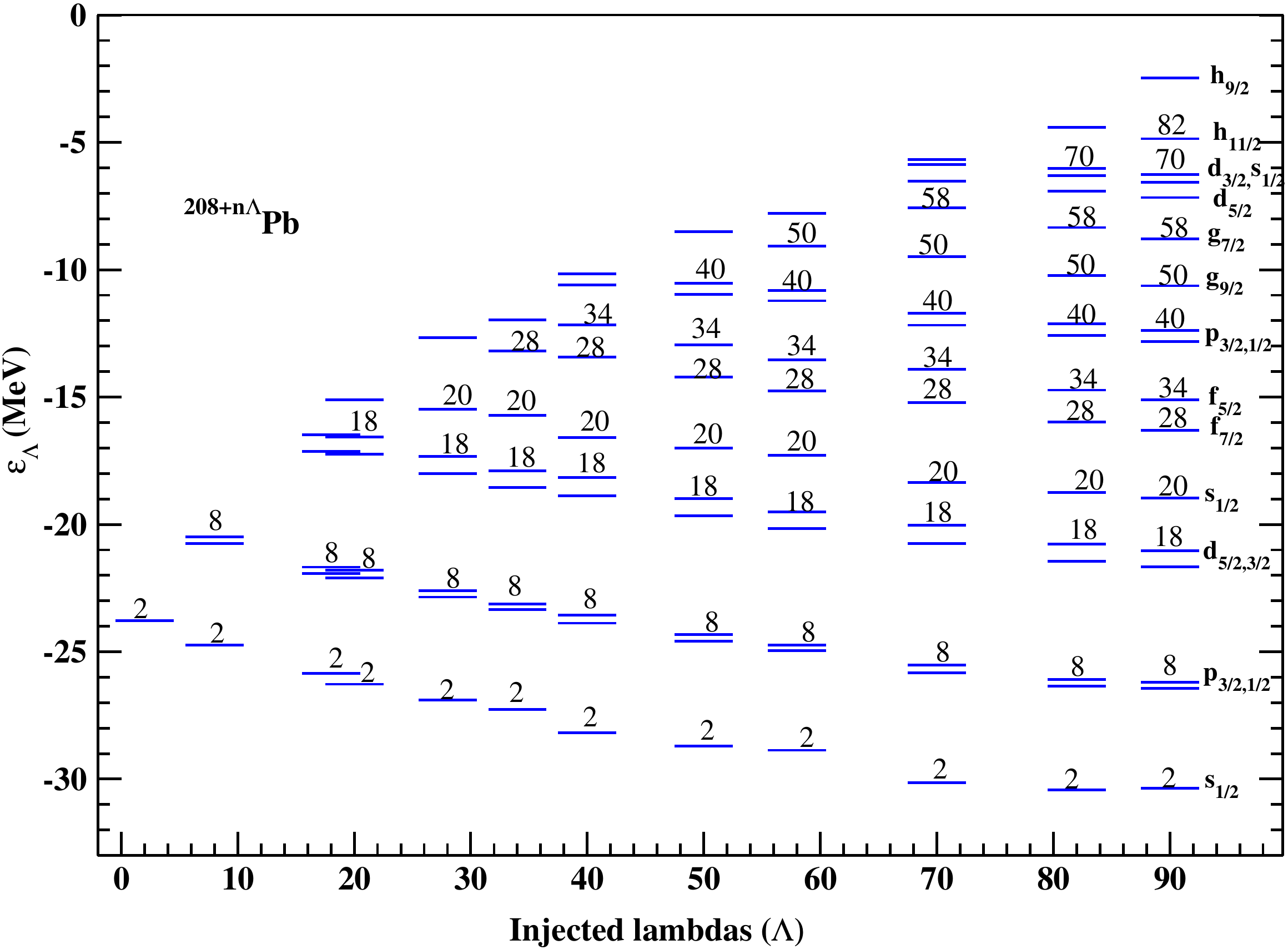}
\caption{\label{levels-pb}(color online) Single-particle energy levels for 
triply magic Pb multi-hypernuclei for 
$\Lambda$ = 2, 8, 18, 20, 28, 34, 40, 50, 58, 70, 82 and 90.}
\end{figure}

\begin{figure}
\includegraphics[width=1.0\columnwidth]{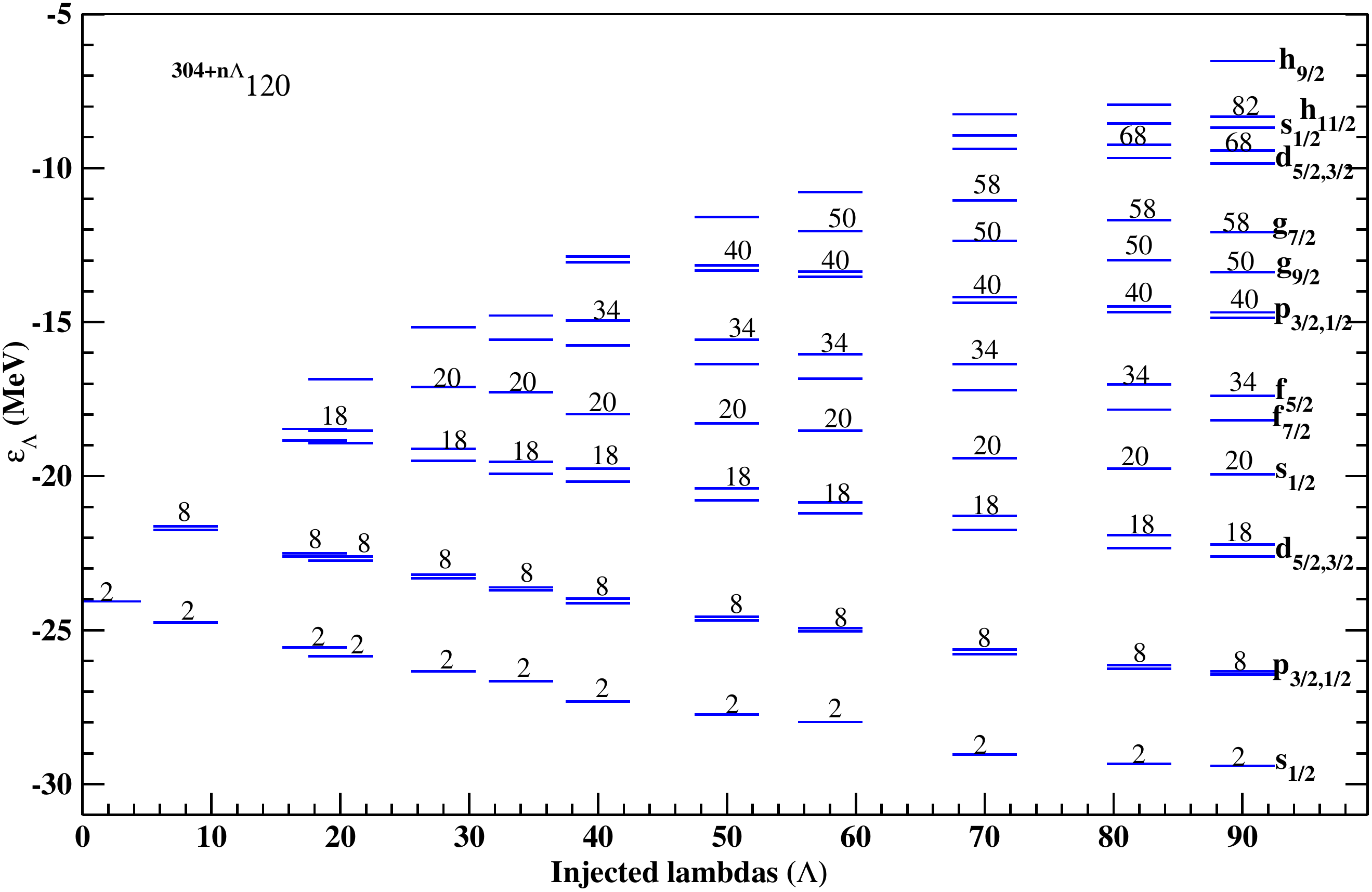}
\caption{\label{levels-120}(color online) Single-particle energy levels for 
triply magic $^{304}$120 multi-hypernuclei for 
$\Lambda$ = 2, 8, 18, 20, 28, 34, 40, 50, 58, 70, 82 and 90.}
\end{figure}

Further, we analyze the lambda single-particle energy levels  for $^{48+n\Lambda}Ca$, 
$^{208+n\Lambda}Pb$ and $^{304+n\Lambda}120$ hypernuclei to extract the lambda 
shell gaps for confirming the $\Lambda$ magic number. 
The lambda energy levels as function of added $\Lambda$'s are given 
in Fig.~\ref{levels-ca},~\ref{levels-pb},~\ref{levels-120}.
The filling of $\Lambda$'s is same as the nucleons following the shell model 
scheme with lambda spin-orbit interaction potential.
It is observed that the single-particle gap of spin-orbit splitting in lambda levels 
is smaller than the nucleons due to weakening strength of lambda spin-orbit interaction.
By analyzing the lambda single-particle energy levels of Ca hypernuclei it is found that 
large energy gap exist in $1d_{5/2}$ to $1d_{3/2}$ or $2s_{1/2}$ and that's why 
lambda magic number 14 is emerged.
Further, $2s_{1/2}$ and $1d_{3/2}$ are very much close to each other 
due to weaker strength of $\Lambda$ spin-orbit interaction.
However, $\Lambda=$ 20 is clearly seen due to large energy gap by $f_{7/2}$ to 
lower orbital.
In case of Pb, the large shell gaps for 
$\Lambda =$ 2, 8, 18, 20, 28, 34, 40, 50, 58, 70 and 82 is appeared.
However, the single-particle gap for lambda number 28 is not so strong as compared to 
others suggesting the feeble magic number. 
The inversion of normal level scheme is noticed and the higher levels fill faster than lower one and 
hence this types of filling is responsible to emerge the new more magic number.
For example, the filling of $1d_{3/2}$ before $2s_{1/2}$ shows a shell gap at $\Lambda=$ 18. 
Along the similar line, due to inversion between \{$1f_{5/2}$, $2p_{3/2}$\} and 
\{$1g_{7/2}$, $2d_{5/2}$\} the $\Lambda$ closed shells 34 and 58 is observed, respectively. 
In case of superheavy multihypernuclei, large single-particle shell gaps are appeared for 
lambda number 2, 8, 18, 20, 34, 40, 50, 58, 70 and 82.
It is quite worth to notice that pronounced energy gaps is noticed 
in $^{208}Pb$ and $^{304}120$ at $\Lambda=$ 34, 58 are being suggested 
to be strong $\Lambda$ shell closure. 
The sharp peaks observed in $\delta_{2\Lambda}$ at $\Lambda=$ 2, 8, 20, 34, 40 and 58 is 
clearly reflected from lambda single-particle energies, where a large energy gap 
is exist for the filling of these number of lambda hyperons.

\begin{table}
\caption{Lambda magic number produced in various considered 
multi-hypernuclei are tabulated here.}
\begin{ruledtabular}
\begin{tabular}{cccccccccccccc}
\multicolumn{1}{c}{Hypernuclei}&\multicolumn{13}{c}{Lambda magic number}\\
\cline{1-1}\cline{2-14}
\hline
$^{16+n\Lambda}$O   &2&8&-&-&-&-&-&-&-&-&-&-&-\\
$^{48+n\Lambda}$Ca  &2&8&14&-&20&-&-&-&-&-&-&-&-\\
$^{58+n\Lambda}$Ni  &2&8&14&-&20&28&34&-&-&-&-&-&-\\
$^{90+n\Lambda}$Zr  &2&8&14&-&20&28&-&40&50&-&-&-&-\\
$^{124+n\Lambda}$Sn &2&8&14&-&20&28&34&40&50&-&-&-&-\\
$^{132+n\Lambda}$Sn &2&8&14&-&20&28&34&40&50&-&-&-&-\\
$^{208+n\Lambda}$Pb &2&8&-&18&20&28&34&40&50&58&-&70&82\\
$^{292+n\Lambda}$120&2&8&-&18&-&28&34&40&50&58&68&-&82\\
$^{304+n\Lambda}$120&2&8&-&18&20&28&34&40&50&58&68&70&82\\
$^{378+n\Lambda}$120&2&8&14&18&20&-&34&40&50&58&68&-&82\\
\end{tabular}
\end{ruledtabular}
\label{tab2}
\end{table}

\subsection{Magicity}
Various signatures of the evolution of magic shell gaps have been discovered 
across the nuclear landscape during the past few decades~\cite{zhang2005,patra2012} such as 
(i) A large binding energy then neighbouring nuclides, 
(ii) Sudden fall at separation energy, (iii) A large shell gap, etc.
It becomes therefore quite relevant to extend the prediction of magic 
numbers to the hypernuclear chart. 
Looking for the magic behaviour firstly, we emphasize on binding of some 
selected nuclei whose nucleonic core is doubly magic such 
as $O$, $Ca$, $Ni$, $Zr$, $Sn$, $Pb$ and $120$.
Some certain number of $\Lambda$ hyperon binds the nuclear core
with maximum stability that may correspond to $\Lambda$ magic number in hypernuclei 
and might form a triple magic system with doubly magic nuclear core as initially 
discussed in Ref.~\cite{zofka1989} and recently in Ref.~\cite{ekhan2015}. 
For example, $\Lambda=$ 2, produce a maximum binding for 
$^{16}O$ and $^{48}Ca$ reveal the maximum binding with $\Lambda=$ 8. 
Also, $^{378}120$ shows a peak binding on addition of 90 $\Lambda$'s.
In this way, we extract some lambda number which are 
2, 8, 18, 20, 40, 70, 90 produce a maximum stability 
for their particular system and suppose to be $\Lambda$ magic number 
as much as close to harmonic oscillator number.
But several other strong signatures are exist to identify the magicity and we make the 
analysis in this direction to look out the correct $\Lambda$ magic number.
After analyzing the $S_\Lambda$ and $S_{2\Lambda}$, for considered 
light to superheavy mass multi-hypernuclei it is 
noticed that a sudden fall is observed at 
$\Lambda=$ 2, 8, 14, 20, 28, 34, 40, 50, 58, 68, 70 and 82.
And hence, the analysis suggest that these numbers are supposed to be 
$\Lambda$ magic number in multi-lambda hypernuclei 
and form a triply magic system having doubly magic nucleonic core.

In order to identify the $\Lambda$ magic number strongly, two-lambda 
shell gaps is examined which provide more strong signature of magicity 
and favoured $S_{2\Lambda}$.
Pronounced peak in two-lambda shell gap is observed at 
$\Lambda=$ 2, 8, 20 and 40 indicating a strong shell closure.
The peaks observed with a significant magnitude at 
$\Lambda=$ 14, 18, 34, 50, 58, 70 and 82 indicating a shell closure also.
Further to testify, we look for the lambda shell gaps by examining the single-particle energy levels. 
We noticed that a large single-particle gap is appeared in $^{208}Pb$ hypernucleus at 
$\Lambda=$ 2, 8, 18, 20, 28, 34, 40, 50, 58 and 70 confirming the lambda magic number. 
The analysis of single-particle energy levels for $^{378}120$ 
multi hypernuclei clear the results by showing a large energy gap 
in 2, 8, 18, 20, 34, 40, 50, 58, 70 and 82.
It is to mention that a significant shell gap is observed for 34 and 58 
suggesting a strong $\Lambda$ shell closure.
The inversion of normal level scheme is responsible to 
emerge the $\Lambda$ magic number 34 and 58. 
The experimentally confirmation of nucleonic shell closure of 34 
supports our predictions~\cite{stepp2013,maierbeck2009}.
The nucleonic number 14, 16, 18 and 32 have also been in discussion 
and expected to be shell closure~\cite{jia2016,ozawa2000,kanungo2002,rodriguez2007}.
In addition, nucleon number 16 and 32 have also been experimentally confirmed in exotic nuclei as 
a new neutron magic number~\cite{kanungo2009,wienholtz2013}.
Therefore, it is concluded that the $\Lambda$ magicity quite resembles 
with the nuclear magicity and it is expected that our predictions 
might be used as significant input to make the things clear regarding new sub shell closure.
The predicted $\Lambda$ magic number in multi-hypernuclei are framed in Table~\ref{tab2}. 
The present lambda magic numbers are quite agreeable with the 
prediction of Ref.~\cite{zofka1989} and Ref.~\cite{ekhan2015} where 
Bruckner-Hartee Fock calculations using the lambda 
density functional have been made. 
It is clear from the plot of $\delta_{2\Lambda}$, that 34, 58 and 70 have peaks 
of great magnitude, while 68 in a feeble 
magnitude and suppose to be subshell closure.
Moreover, strong nucleonic magic number 28 is observed very feeble in lambda magicity.
It is to mention that the lambda number 14 is appeared in medium mass 
hypernuclei, on contrary to this 18 is observed in superheavy mass multi hypernuclei.

\section{Summary and conclusion}\label{sum}
In summary, we have suggested the possible $\Lambda$ magic 
number i.e. 2, 8, 14, 18, 20, 28, 34, 40, 50, 58, 68, 70, 82 in 
multi-$\Lambda$ hypernuclei within the relativistic mean field theory with 
effective $\Lambda$N as well as $\Lambda$$\Lambda$ interactions.
The survey of $\Lambda$ magic number is made on the basis of binding energy, 
one- and two-lambda separation energies $S_\Lambda$, $S_{2\Lambda}$, and 
two-lambda shell gaps $\delta_{2\Lambda}$. 
It is noticed that pronounced single particle energy gap is observed for 
lambda number 34 and 58 in Pb and superheavy multi hypernuclei 
representing the strong $\Lambda$ magic number.
Our predictions are strongly supported by nuclear magicity, where nucleon number 34 is experimentally 
confirmed as a neutron shell closure~\cite{stepp2013,maierbeck2009}.
It is expected that the weakening strength of lambda spin-orbit interaction 
potential is responsible for emerging the new lambda magic number.
The predicted $\Lambda$ magic numbers are in remarkable agreement with earlier 
predictions~\cite{ekhan2015,zofka1989} and hypernucler magicity quite resembles with nuclear magicity.
In the analogy of nuclear stability, we noticed a similar pattern of binding 
energy per particle in hypernuclear regime and Ni hypernucleus with 8 $\Lambda$'s is 
found to be most tightly bound triply magic system in hypernuclear landscape.
The addition of $\Lambda$ hyperons have significant impact on nucleon distribution 
and remove the bubble structure partially or fully.
The spin-orbit interaction potentials and mean field potentials is also 
studied for predicted triply magic hypernuclear systems and the added $\Lambda$'s 
affect both the potential to a large extent. 
The present results may be used as a significant input to produce the 
triply magic hypernuclei in laboratory in future. 
It is also concluded that the addition of $\Lambda$ hyperons draw the nuclear 
system towards more stability with increasing strangeness. 
We noticed that the core of superheavy nuclei has more affinity 
to absorb large number of hyperons. 
This means such systems are able to simulate the strange hadronic matter 
containing large number of heavy hyperons such as $\Sigma$'s and $\Xi$'s including 
several $\Lambda$'s and the formation of such systems has large implication in nuclear-astrophysics. 

\section{Acknowledgments}
One of the authors (MI) acknowledge the hospitality 
provided by Institute of Physics, Bhubaneswar during the work.



\begin{thebibliography}{999}
\bibitem{bando1990}
H. Bando, T. Motoba and J. Zofka, Int. J. Mod. Phys. A {\bf5}, 4021 (1990).
\bibitem{tan2004}
Y. Tan, X. Zhong, C. Cai, P. Ning, Phys. Rev. C {\bf70}, 054306 (2004).
\bibitem{lu2011}
B. Lu, E. Zhao and S. Zhou, Phys. Rev. C {\bf84}, 014328 (2011).
\bibitem{3}
M. Rayet, Ann. Phys. {\bf102}, 226 (1976).
\bibitem{4}
M. Rayet, Nucl. Phys. A {\bf367}, 381 (1981).
\bibitem{5}
D. E. Lanskoy, Phys. Rev. C {\bf58}, 3351(1998).
\bibitem{6}
J. Meng, H. Toki, S. G. Zhou et.al., Prog. Part. Nucl. Phys. {\bf57}, 470 (2006).
\bibitem{7}
X.-R. Zhou, H.-J. Schulze, H. Sagawa, C.-X. Wu and E.-G. Zhao, Phys. Rev. C {\bf76}, 034312 (2007).
\bibitem{8}
H. F. L\"u, Chin. Phys. Lett. {\bf25}, 3613 (2008).
\bibitem{9}
N. Guleria, S. K. Dhiman and R. Shyam, Nucl. Phys. A {\bf886}, 71 (2012).
\bibitem{10}
F. Minto and K. Hagino, Phys. Rev. C {\bf85}, 024316 (2012).
\bibitem{11}
A. Bouyssy, Nucl. Phys. A {\bf381}, 445 (1982).
\bibitem{14}
J. Mares and J. Zofka, Z. Phys. A {\bf333}, 209 (1989).
\bibitem{15}
M. Rufa, J. Schaffner, J. Maruhn H. St\"ocker and W. Griner and P.-G. Reinhard, Phys. Rev. C. {\bf42}, 2469(1990).
\bibitem{16}
J.  Schaffner, C. Griner and H. St"ocker, Phys. Rev. C. {\bf46}, 322 (1992).
\bibitem{17}
J. Mares and J. Zofka, Z. Phys. A {\bf345}, 47 (1993).
\bibitem{18}
J. Schaffner, et. al., Ann. Phys. {\bf235}, 35 (1994).
\bibitem{19}
E. N. E. Van Dalen, G. Colucci and A. D. Sedrakian, Phys. Lett. B {\bf734}, 383 (2014). 
\bibitem{Hiyama2010}
E. Hiyama, M. Kamimura, Y. Yamamoto, T. Motoba and T. A. Rijken, Prog, Theo. Phys. Suppl. {\bf185}, 106 (2010).
\bibitem{Hiyama2012} 
E. Hiyama, Few. Body. Syst. {\bf53}, 189 (2012).
\bibitem{Nemura2005} 
H. Nemura, S. Shinmura, Y. Akaishi, and Khin Swe Myint, Phys. Rev. Lett. {\bf94}, 202502 (2005).
\bibitem{nemura2005}
H. Nemura, S. Shinmura, Y. Akaishi, and Khin Swe Myint, Nucl. Phys. A {\bf754}, 110c (2005).
\bibitem{Gal2009} 
A. Gal, Few. Body. Syst. {\bf45}, 105 (2009).
\bibitem{Gal2011} 
A. Gal and D. J. Millener, Phys. Lett. B {\bf701}, 342 (2011).
\bibitem{Usmani2006} 
A. A. Usmani, Phys. Rev. C {\bf73}, 011302(R) (2006).
\bibitem{Usmani2008} 
A. A. Usmani and F. C. Khanna, J. Phys. G: Nucl. Part. Phys. {\bf35}, 025105 (2008).
\bibitem{Zhou2008} 
X.-R. Zhou, A. Polls, H. -J. Schulze, and I. Vida\~na, Phys. Rev. C {\bf78}, 054306 (2008).
\bibitem{Vidana2004} 
I. Vida\~na, A. Ramos and A. Polls, Phys. Rev. C {\bf70}, 024306 (2004).
\bibitem{Samanta2008} 
C. Samanta, P. Roy Chowfhury and D. N. Basu, J. Phys. G: Nucl. and Part. Phys. {\bf35}, 065101 (2008).
\bibitem{25}
D. E. Lenskoy and Y. Yamamoto, Phys. Rev. C {\bf55}, 2330 (1997).
\bibitem{26}
J. Cugnon, A. Lejeune and H.-J. Schulze, Phys. Rev. C {\bf62}, 064308 (2000).
\bibitem{27}
I. Vidana, A. Polls, A. Ramos and H.-J. Schulze, Phys. Rev. C {\bf64}, 044301 (2001).
\bibitem{goeppert1958}
M. Goeppert-Mayer and J. Jensen, Elementary Theory of Nuclear Shell Structure  
(Wiley, New York, 1955; Inostrannaya Literatura, Moscow, 1958).
\bibitem{solovev1992}
V. G. Solov’ev, Theory of Atomic Nuclei  (Energoizdat,
Moscow, 1981; Institute of Physics, Bristol, England, 1992).
\bibitem{nilsson1995}
S. G. Nilsson and I. Ragnarsson, Shapes and Shells in Nuclear Structure Cambridge University Press, Cambridge, England, (1995).
\bibitem{sorlin2008}
O. Sorlin, M.-G. Porquet, arXive:0805.2561v1 (2008).
\bibitem{patra2012}
M. Bhuyan and S. K. Patra, Mod. Phys. Lett. A {\bf27}, 1250173 (2012).
\bibitem{zhang2005}
W. Zhang, J. Meng., S. Q. Zhang, L.S. Geng and H. Toki, Nucl. Phys. A {\bf753}, 106 (2005).
\bibitem{strutinsky1967} 
V. M. Strutinsky, Nucl. Phys. A {\bf95}, 420 (1967).
\bibitem{strutinsky1998} 
V. M. Strutinsky, Nucl. Phys. A {\bf122}, 1 (1998).
\bibitem{patra1997}
S. K. Patra, R. K. Gupta and W. Greiner, Mod. Phys. Lett. A {\bf12}, 1727 (1997).
\bibitem{patra2004}
T. Sil, S.K. Patra, B.K. Sharma, M. Centelles, and X. Vinas, Phys. Rev. C {\bf69}, 044315 (2004).
\bibitem{rutz1997}
K. Rutz, M. Bender, T. B\"urvenich, T. Schilling, P.-G. Reinhardt, J. A. Maruhn, and W. Greiner, Phys. Rev. C {\bf56}, 238 (1997).
\bibitem{patra1999}
S. K. Patra, C.-L. Wu, C. R. Praharaj, and R. K. Gupta, Nucl. Phys. A {\bf651}, 117 (1999).
\bibitem{meng2007}
J. Meng, H. Toki, S. Zhou, S. Zhang, W. Long, and L. Geng, Prog. Part. Nucl. Phys. {\bf57}, 470 (2007).
\bibitem{ikram2013}
S. K. Singh, M. Ikram and S. K. Patra, Int. J. Mod. Phys. E {\bf22}, 1350001 (2013).
\bibitem{serot1992}
B. D. Serot, Rep. Prog. Phys. {\bf55}, 1855 (1992).
\bibitem{gambhir1990}
Y. K. Gambhir, P. Ring and A. Thimet, Ann. Phys. (NY) {\bf198}, 132 (1990).
\bibitem{ring1996}
P. Ring, Prog. Part. Nucl. Phys. {\bf37}, 193 (1996).
\bibitem{serot1986}
B. D. Serot and J. D. Walecka, Adv. Nucl. Phys. {\bf16}, 1 (1986).
\bibitem{boguta1977}
J. Boguta and A. R. Bodmer, Nucl. Phys. A {\bf292}, 413 (1977).
\bibitem{jha2007}
T. K. Jha, P. K. Raina, P. K. Panda and S. K. Patra, Phys. Rev. C {\bf74}, 029903 (2007).
\bibitem{glendenning1998}
N. K. Glendenning, J. Schaffner, Phys. Rev. Lett {\bf81}, 4564 (1998).
\bibitem{schaffner2002}
J. Schaffner, M. Hanauske, H. St\"ocker and W. Greiner, Phys. Rev. Lett. {\bf89}, 171101 (2002).
\bibitem{sugahara1994}
Y. Sugahara and H. Toki, Prog. Theor. Phys. {\bf92}, 803 (1994).
\bibitem{vretenar1998}
D. Vretenar, W. Po\"schl, G. A. Lalazissis and P. Ring, Phys. Rev. C {\bf57}, R1060 (1998).
\bibitem{lu2003}
H. -F. L\"u, J. Meng, S. Q. Zhang and S. G. Zhou, Eur. Phys. J. A. {\bf17}, 19 (2003).
\bibitem{shen2006}
H. Shen, F. Yang and H. Toki, Prog. Theor. Phys. {\bf115}, 325 (2006).
\bibitem{win2008}
M. T. Win and K. Hagino, Phys. Rev. C {\bf78}, 054311 (2008).
\bibitem{ikram14}
M. Ikram, S. K. Singh, A. A. Usmani and S. K. Patra, Int. J. Mod. Phys. E {\bf23}, 1450052 (2014).
\bibitem{schaffner1994} 
J. Schaffner, C. B. Dover, A. Gal, C. Greiner, D. J.
Millener and H. St\"ocker, Ann. Phys. (N.Y.) {\bf235}, 35 (1994).
\bibitem{schaffner1993}
J. Schaffner, C. B. Dover, A. Gal, C. Greiner and H. St\"ocker, Phys. 
Rev. Lett. {\bf71}, 1328 (1993).
\bibitem{glendenning1993}
N. K. Glendenning, D. Von-Eiff, M. Haft, H. Lenske and M. K. Weigel, Phys. Rev. C {\bf48}, 889 (1993).
\bibitem{mares1994}
J. Mares and B. K. Jennings, Phys. Rev. C {\bf49}, 2472 (1994).
\bibitem{rufa1990}
M. Rufa, J. Schaffner, J. Maruhn, H. St\"ocker and W. Greiner, Phys. Rev. C {\bf42}, 2469 (1990).
\bibitem{lalazissis09}
G. A. Lalazissis, S. Karatzikos, R. Fossion, D. Pena Arteaga, A. V. Afanasjev and P. Ring, Phys. Lett. B {\bf671}, 36 (2009). 
\bibitem{chiapparini09}
M. Chiapparini, M. E. Bracco, A. Delfino, M. Malheiro, D. P. Menezes, C. Providencia, Nucl. Phys. A {\bf826}, 178 (2009).
\bibitem{dover1984}
C. B. Dover and A. Gal, Prog. Part. Nucl. Phys. {\bf12}, 171 (1984).
\bibitem{millener1988}
D. J. Millener, C. B. Dover and A. Gal, Phys. Rev. C {\bf38}, 2700 (1988).
\bibitem{hashimoto2006}
O. Hashimoto and H. Tamura, Prog. Part. Nucl. Phys. {\bf57}, 564 (2006).
\bibitem{ekhan08}
E. Khan et.al, Nucl. Phys. A {\bf800}, 37 (2008).
\bibitem{decharge03}
J. Decharge, et., al. Nucl. Phys. A {\bf716}, 55 (2003).
\bibitem{grasso2009}
M. Grasso et. al., Phys. Rev. C {\bf79}, 034318 (2009).
\bibitem{wang2013}
Y. Z. Wang, J. Z. Gu, Z. Y. Li, and Z. Y. Hou, Eur. Phys. J. A {\bf15}, 49 (2013).
\bibitem{nakada2013}
H. Nakada, K. Sugimura and J. Margureon, Phys. Rev. C {\bf87}, 067305 (2013).
\bibitem{yao2012}
J. M. Yao, H. Mei and Z. P. Li, Phys. Lett. B {\bf723}, 459 (2013).
\bibitem{yao2013}
J. M. Yao, S. Baroni, M. Bender and P.-H. Heenen, Phys. Rev. C {\bf86}, 014310 (2012).
\bibitem{wu2014}
X. Y. Wu, T. M. Yao, and Z. P. Li, Phys. Rev. C {\bf89}, 017304 (2014).
\bibitem{ekhan2015}
E. Khan, J. Margueron, F. Gulminelli and Ad. R. Raduta, Phys. Rev. C {\bf92}, 044313 (2015).
\bibitem{boguta1981}
J. Boguta and S. Bohrmann, Phys. Lett. B {\bf102}, 93 (1981).
\bibitem{noble1980}
J. V. Noble, Phys. Lett. B {\bf89}, 325 (1980).
\bibitem{brockmann1977}
R. Brockmann and W. Weise, Phys. Lett. B {\bf69}, 167 (1977).
\bibitem{ajimura2001}
S. Ajimura et. al., Phys. Rev. Lett. {\bf86}, 4255 (2001).
\bibitem{keil2000}
C. M. Keil, F. Hoffmann and H. Lenske, Phys. Rev. C {\bf61}, 064309 (2000).
\bibitem{zofka1989}
J. Mares and J. Zofka, Z. Phys. A. {\bf333}, 209 (1989).
\bibitem{stepp2013}
D. Steppenbeck, S. Takeuchi, N. Aoi, P.Doornenbal, M. Matsushita et.al., Nature {\bf502}, 207 (2013).
\bibitem{maierbeck2009}
P. Maierbeck et. al., Phys. Lett. B {\bf675}, 22 (2009).
\bibitem{rodriguez2007}
T. R. Rodriguez, J. L. Egido, Phys. Rev. Lett. {\bf99}, 062501 (2007).
\bibitem{ozawa2000}
A. Ozawa et. al., Phys. Rev. Lett. {\bf84}, 5493 (2000).
\bibitem{kanungo2002}
R. Kanungo et.al., Phys. Lett. B {\bf528}, 58 (2002).
\bibitem{jia2016}
Jia. Jie Li, J. Margueron, W. H. Long, N. V. Giai, Phys. Lett. B {\bf753}, 97 (2016).
\bibitem{kanungo2009}
R. Kanungo et. al., Phys. Rev. Lett. {\bf102}, 152501 (2009).
\bibitem{wienholtz2013}
F. Wienholtz et. al., Nature {\bf498}, 346 (2013).
\end{thebibliography}
\end{document}